\documentclass[lettersize,journal]{IEEEtran}
\usepackage{amsmath,amsfonts}
\usepackage[ruled, linesnumbered]{algorithm2e}
\usepackage{algorithmic}
\usepackage{bm}
\usepackage{array}
\usepackage[caption=false,font=normalsize,labelfont=sf,textfont=sf]{subfig}
\usepackage{textcomp}
\usepackage{url}
\usepackage{verbatim}
\usepackage{graphicx}
\usepackage{cite}
\usepackage{comment}
\usepackage{multicol}
\usepackage{multirow}
\usepackage{colortbl}
\usepackage[caption=false]{subfig}
\usepackage{booktabs}
\usepackage{hyperref}
\usepackage{tikz}
\usepackage{balance}
\usepackage{amsmath,amssymb,amsfonts}
\usepackage{subcaption}
\usepackage{footmisc}
\usepackage{hyperref}
\usepackage{caption}
\usepackage{comment}
\usepackage{soul}
\usepackage{dblfloatfix}

\begin{document}

\title{Echocardiography to Cardiac MRI View Transformation for Real-Time Blind Restoration}

\author{Ilke Adalioglu, Serkan Kiranyaz, Mete Ahishali, Aysen Degerli, Tahir Hamid, Rahmat Ghaffar, Ridha Hamila, and Moncef Gabbouj
\thanks{This work was partly supported by the NSF CBL Program under Project AMaLIA funded by Business Finland.}
\thanks{Ilke Adalioglu, Mete Ahishali, and Moncef Gabbouj are with the Faculty of Information Technology and Communication Sciences, Tampere University, Tampere, Finland (email: \textit{ilke.adalioglu@tuni.fi, mete.ahishali@tuni.fi, moncef.gabbouj@tuni.fi}).}
\thanks{Serkan Kiranyaz and Ridha Hamila are with the Department of Electrical Engineering, Qatar University, Doha, Qatar (email: \textit{mkiranyaz@qu.edu.qa, hamila@qu.edu.qa}).}
\thanks{Aysen Degerli is with VTT Technical Research Centre of Finland, Tampere, Finland (email: \textit{aysen.degerli@vtt.fi}).}
\thanks{Tahir Hamid is with the Hamad Medical Corporation Hospital, Doha, Qatar (email:\textit{ tahirhamid76@yahoo.co.uk}}  
\thanks{Rahmat Ghaffar is with the Hayatabad Medical Complex, Peshawar, Pakistan (email:\textit{rahmat.cardio@gmail.com})}
}
\maketitle




\maketitle

\begin{abstract}

 Echocardiography is the most widely used imaging to monitor cardiac functions, serving as the first line in early detection of myocardial ischemia and infarction. However, echocardiography often suffers from several artifacts including sensor noise, lack of contrast, severe saturation, and missing myocardial segments which severely limit its usage in clinical diagnosis. In recent years, several machine learning methods have been proposed to improve echocardiography views. Yet, these methods usually address only a specific problem (e.g. denoising) and thus cannot provide a robust and reliable restoration in general. On the other hand, cardiac MRI provides a clean view of the heart without suffering such severe issues. However, due to its significantly higher cost, it is often only afforded by a few major hospitals, hence hindering its use and accessibility. 
In this pilot study, we propose a novel approach to transform echocardiography into the cardiac MRI view. For this purpose, \textit{Echo2MRI} dataset, consisting of echocardiography and real cardiac MRI image pairs, is composed and will be shared publicly. A dedicated Cycle-consistent Generative Adversarial Network (Cycle-GAN) is trained to learn the transformation from echocardiography frames to cardiac MRI views.     
An extensive set of qualitative evaluations shows that the proposed transformer can synthesize high-quality artifact-free synthetic cardiac MRI views from a given sequence of echocardiography frames. Medical evaluations performed by a group of cardiologists further demonstrate that synthetic MRI views are indistinguishable from their original counterparts and are preferred over their initial sequence of echocardiography frames for diagnosis in \textbf{78.9\%} of the cases. 
\end{abstract}

\begin{IEEEkeywords}
Blind Echocardiography Restoration, Echocardiography to MRI View Transformation, Generative Adversarial Networks, Magnetic Resonance Imaging, Virtual MRI Machine 
\end{IEEEkeywords}

\section{Introduction}
\label{sec:introduction}

\IEEEPARstart{C}{ardiovascular} diseases are responsible for 32\% of global mortality. Hence, identifying cardiovascular health issues at early stages is crucial in reducing mortality and morbidity by offering early treatment and ultimately improving the patient outcome\cite{WHO}. As an essential tool, cardiac imaging visualizes cardiac structure and its function, playing a crucial role in the early detection of cardiac disorders such as coronary artery disease (CAD)
\cite{counseller2023recent, dalen2014CAD}.

\begin{figure}[t]
    \centering
    \includegraphics[width=\linewidth]{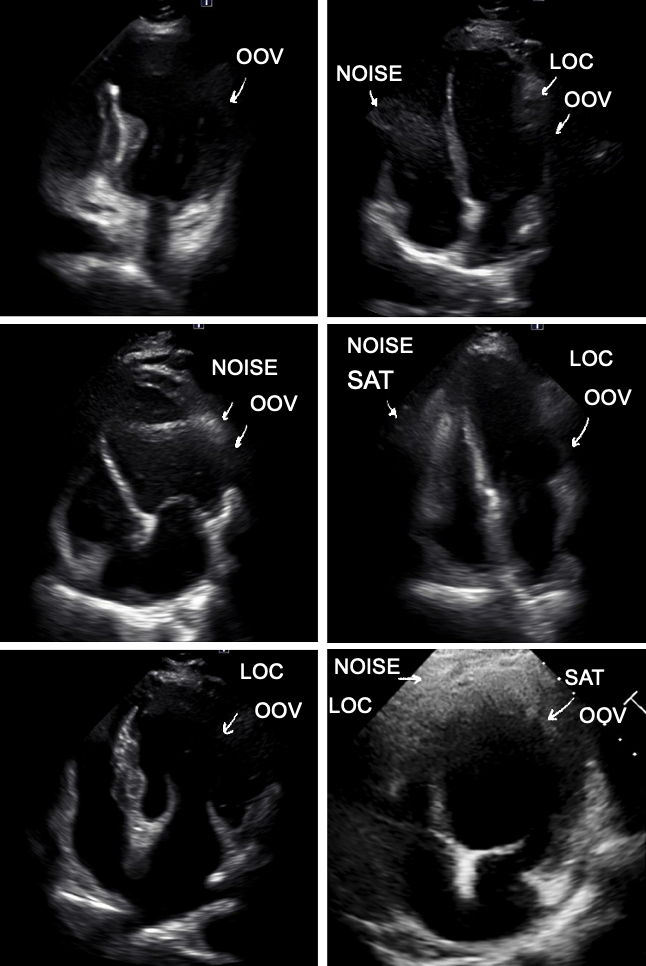}
    \caption{Sample echocardiography frames from the Echo2MRI dataset demonstrating poor quality due to common artifacts such as saturation (SAT), out-of-view (OOV), noise, and lack of contrast (LOC).}
    \label{fig:issues}
\end{figure}

Echocardiography is the most accessible cardiac imaging modality due to its low cost and portability. Therefore, it has been widely used even in resource-limited environments \cite{pennell2004_5_5xcost, bullock2024multimodality}. Cost efficiency and mobility of echocardiography devices are substantially important since over three-quarters of CAD-related fatalities occur in low and middle-income countries \cite{WHO}. 
However, due to its acoustic nature, echocardiography suffers from echogenicity artifacts, where undesired echoes and distortions degrade image quality \cite{kurtz1981echogenicity}. This blend of artifacts includes speckle noise, a granular pattern caused by scattering of the echoes, which results in a loss of clarity, and discontinuation of structure boundaries hindering the overall visibility \cite{wu2015echocardiogram|_multiple_sources_of_noise_data_driven_denoising}. Other compound artifacts in the sensory acquisition may result in images lacking contrast and saturation within the limited acoustic window, causing certain regions of the heart to be unrecognizable as illustrated in Fig. \ref{fig:issues}. Since the movement of the left ventricular (LV) walls provides crucial diagnostic information such as the identification of the regional wall motion abnormalities (RWMA) \cite{kinno2017comparison, votavova2015LVecho}, it becomes challenging and even impossible for cardiologists to perform a reliable diagnosis when such compound artifacts distort the movement of the LV wall. Worst of all, some parts of the LV wall may often fall outside the captured view, making any RWMA analysis entirely infeasible.

Many approaches have been developed \cite{diller2019denoising, wu2015echocardiogram|_multiple_sources_of_noise_data_driven_denoising, khare2010wavelet_denoising, gupta2004wavelet} for removing speckle noise in echocardiography. However, these methods are often ineffective because speckle noise is only one of many artifacts corrupting the frames while other artifacts including broader issues of obscured regions, lack of contrast, and severe saturation caused by the echoed noise still survive any denoising operation. As a result, such traditional denoising methods can provide no or minimal restoration performance for such a random blend of artifacts in a typical echocardiography. A robust and reliable restoration solution should, therefore, be ``blind" and restore the input without any prior knowledge about the type and severity of the artifacts. 
To effectively perform a blind restoration, we focus on innovative approaches beyond denoising by incorporating methods to reconstruct out-of-view regions and obscured parts due to, e.g., low contrast and high saturation. 

Cardiovascular magnetic resonance imaging (cardiac MRI) is the gold standard for the assessment of myocardial structure and its function \cite{collins2015globalCMRgold, salerno2017gold_standards, bruder2005detection}. Thanks to the recent advancements in acquisition speed, and temporal and spatial resolution, cardiac MRI imaging enables highly accurate diagnostic capabilities, solidifying its role as a critical tool in cardiology. However, due to the high cost of equipment and maintenance, a cardiac MRI acquisition is approximately 5.5 times more expensive for patients than echocardiography, making cardiac MRI affordable by only a few major hospitals \cite{pennell2004_5_5xcost}. Accordingly, the estimated accessibility of cardiac MRI is only about 34\% globally, preventing its widespread use \cite{murali2023bringing34}. This disparity highlights the immediate challenge of providing equitable access to high-quality cardiac MRI and improved accessible imaging modality even in remote places, especially in developing countries \cite{bullock2024multimodality}.

As a remedy to the aforementioned drawbacks and limitations of current echocardiography restoration methods, this study proposes a pioneer blind restoration approach that can remove the aforementioned artifacts in typical echocardiography by synthesizing its corresponding cardiac MRI view. Therefore, the proposed approach aims to learn a one-to-one transformation from the 4-chamber echocardiography to the corresponding cardiac MRI view. Such a transformation approach naturally removes the blend of artifacts by performing a modality transformation where the source domain echocardiography is mapped to the target domain cardiac MRI, resulting in a synthetic cardiac MRI view with enhanced contrast and clarity, thus improving the quality and accuracy of diagnostic capabilities.  

\begin{figure}[t]
    \centering
    \begin{tikzpicture}
        \node[anchor=south west, inner sep=0] (image) at (0,0) {\includegraphics[width=0.5\textwidth]{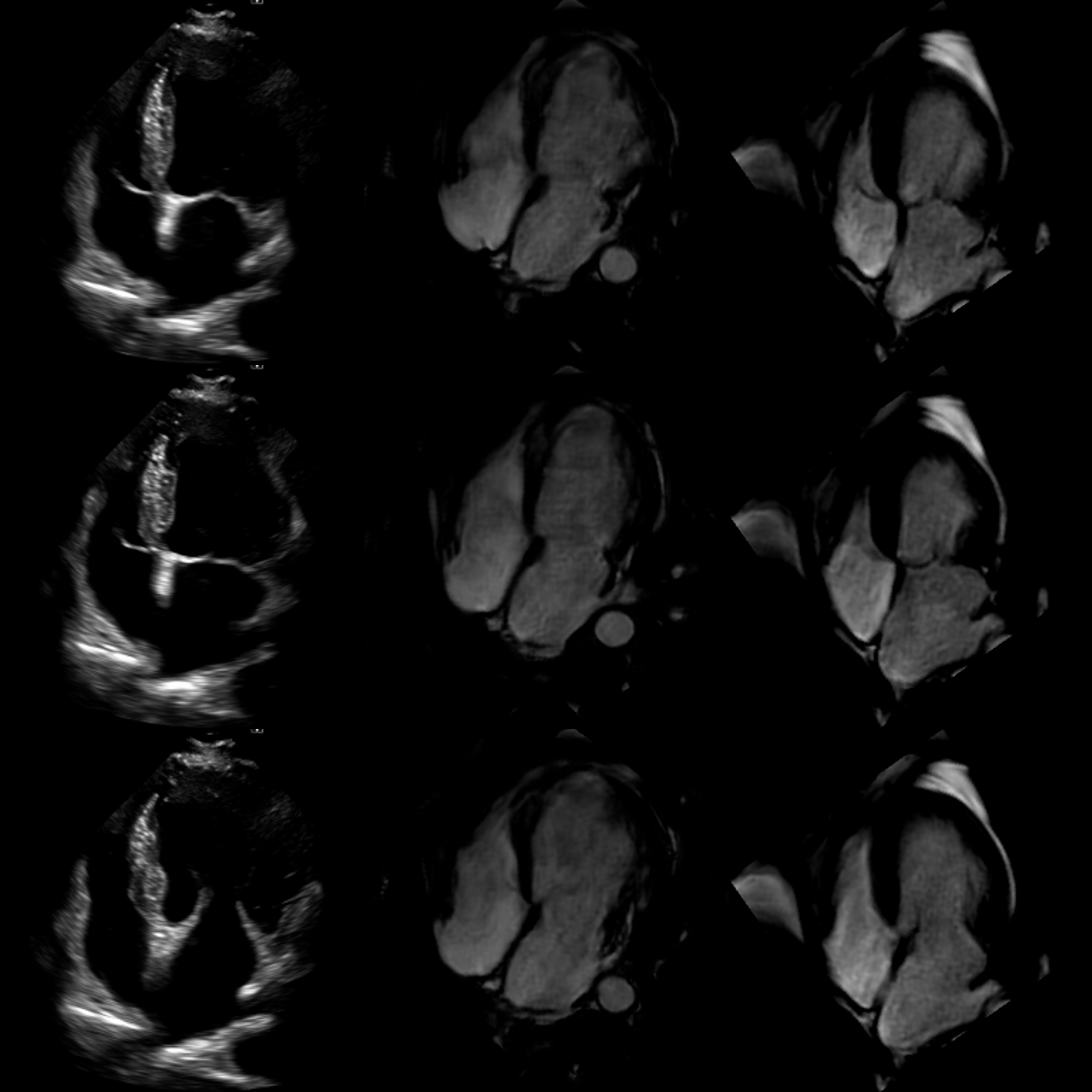}}; 
        \begin{scope}[x={(image.south east)}, y={(image.north west)}]
            \node[anchor=center, text=white] at (0.03, 0.97) {A};
            \node[anchor=center, text=white] at (0.37, 0.97)  {B};
            \node[anchor=center, text=white] at (0.70, 0.97) {C};
        \end{scope}
    \end{tikzpicture}
    \caption{Comparison of 4-chamber Echocardiography (A), generated synthetic cardiac MRI view (B), and the original cardiac MRI view (C) across different phases of the cardiac cycles. The generated synthetic cardiac MRI views successfully interpolate the LV wall's missing segments while preserving the cardiac motion's temporal coherence.}
    \label{Fig: intro_qualitative}
\end{figure}
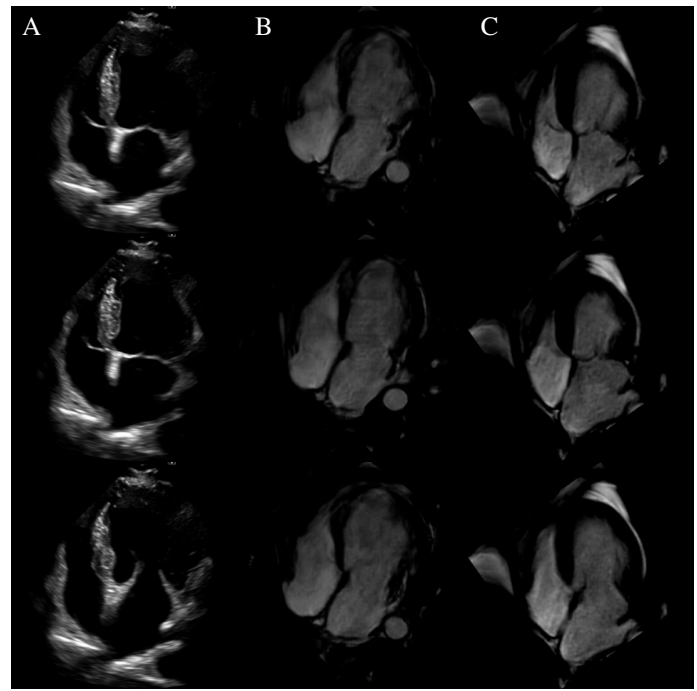

 Fig. \ref{Fig: intro_qualitative} highlights the potential of this study illustrating the reconstructed LV wall in the transformed synthetic cardiac MRI view along with the enhanced visualization of the apex fully restored from the saturation and noise by keeping information consistent. To learn such a blind domain transformation, the proposed approach leverages Cycle-Consistent Generative Adversarial Networks (Cycle-GANs) \cite{zhu2017unpairedCycleGan} which can perform image translation on unpaired data, while preserving the existing patterns in the source domain.

To accomplish this objective, we collected the dataset Echo2MRI consisting of apical $4-$chamber echocardiography and cardiac MRI views of patients from Hayatabad Medical Complex in Pakistan. The multi-view dataset has $52$ patients, of which $36$ have paired views. 

\begin{figure*}[!b]
    \centering
    \includegraphics[width=\linewidth]{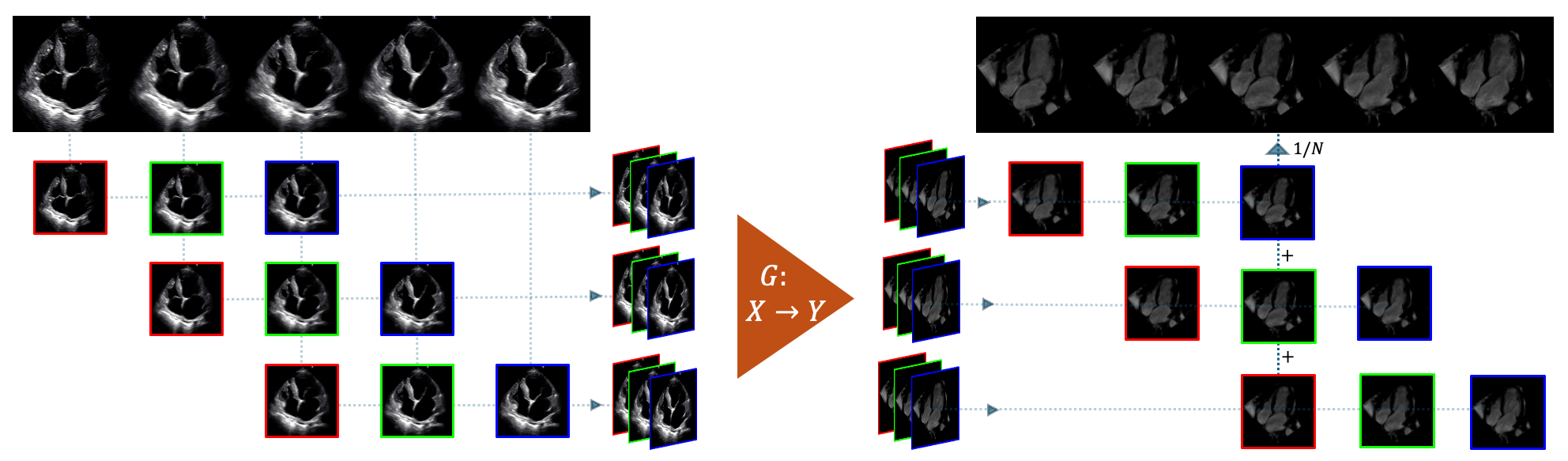}
    \caption{Illustration of pre- and post-processing steps. Preprocessing involes encapsulating temporal information, $t-1$, $t$ and, $t+1$, into a single 3-channel image. After translating from the echocardiography ($X$) domain to the MRI ($Y$) domain with the Generator $G$, the resulting frames also retain temporal information. In the post-processing, output frame is obtained by averaging the slices of timestamps from consecutive frames.}
    \label{fig: post-process}
\end{figure*}

To learn the transformation most effectively, both source (echocardiography) and target (cardiac MRI views) input data are first preprocessed and augmented to incorporate temporal information in modality transformation.

The novel contributions of this study can be summarized as follows:
\begin{itemize}
    \item This study proposes a pioneer modality transformation approach from the source domain echocardiography to the target domain cardiac MRI view.

    \item The transformation enables a blind restoration against any random blend of artifacts typically corrupting the echocardiography. 
    
    \item We compose the first dataset, Echo2MRI, encapsulating paired echocardiogram and cardiac MRI views from the same patients and will make it publicly available for the research community. 

    \item An extensive set of experimental evaluations demonstrates that the proposed 
    transformed synthetic cardiac MRI view improves diagnosis by the cardiologist due to the superior quality and clarity achieved.

    \item Overall, this pilot study paves the way to create the first virtual cardiac MRI machine, thus presenting a great potential for improving cardiac healthcare with significantly reduced costs. 
    
\end{itemize}

The rest of the paper is organized as follows:  Section \ref{sec: method} presents the proposed methodology for echocardiography restoration using modality transformation. Next, the composed Echo2MRI dataset and experimental results are detailed in Section \ref{sec: experimental}. Finally, Section \ref{sec: conclusion} concludes this study with a discussion of possible future research directions.

\section{Proposed Methodology}
\label{sec: method}
In this section, we will first explain pre- and post-processing steps over the echocardiography frames to boost the learning performance for an accurate and reliable transformation. Then, we will detail the proposed echocardiography to cardiac MRI transformation using the Cycle-GANs.

\begin{figure*}[t]
    \centering
    \includegraphics[width=\linewidth]{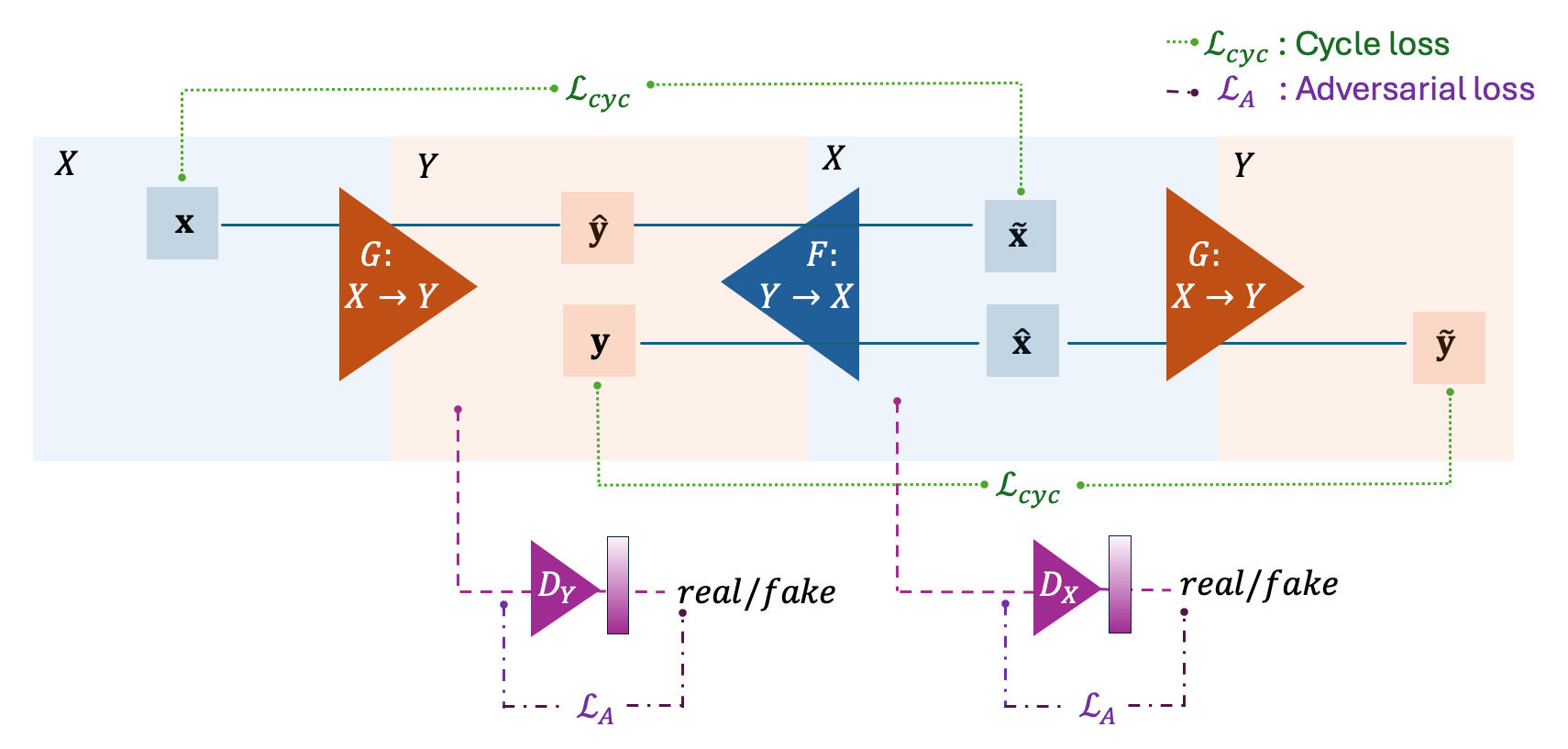}
    \caption{Cycle-GAN architecture for image-to-image translation between Source Domain $X$, and Target Domain $Y$,  corresponding to echocardiography to cardiac MRI. The generator $G:X\rightarrow Y$ translates echocardiography images to the cardiac MRI domain, whereas the generator $F:Y\rightarrow X$ performs the inverse translation from cardiac MRI to echocardiography. Discriminators $D_X$ and $D_Y$ distinguish real and generated images in the respective domains. Losses $\mathcal{L}_{A}$ and $\mathcal{L}_{\text{cyc}}$ are defined in \eqref{eq:gen-cost1} and \eqref{eq:cyc-cost} respectively.}
    \label{fig:CGAN}
\end{figure*}

Initially, we started by transforming each echocardiography frame independently to achieve domain translation frame sequence. However, this method is prone to process-induced artifacts, resulting in undesirable flickering effects. To mitigate these issues and enhance the output quality, we augment the input with temporal information across frames before the translation to reduce flickering and improve the reliability of the output.
Both input and output data are 3-channel images, where the channels represent consecutive time frames: $t-1$, $t$, and $t+1$, represented as red, green, and blue respectively, as shown in Fig. \ref{fig: post-process}. After the transformation, the post-processing step consists of averaging timestamps from consecutive frames. The first and last frames are used without averaging, while the second frame from the beginning and the end are formed by averaging only the two consecutive frames available. It is worth mentioning that incorporating temporal information from distinct transformation instances significantly reduces flickering, especially at the boundaries of the LV wall, and thus, ensures coherent motion across frames, leading to smoother transitions. 

\begin{figure}[!b]
    \centering
    \includegraphics[width=\linewidth]{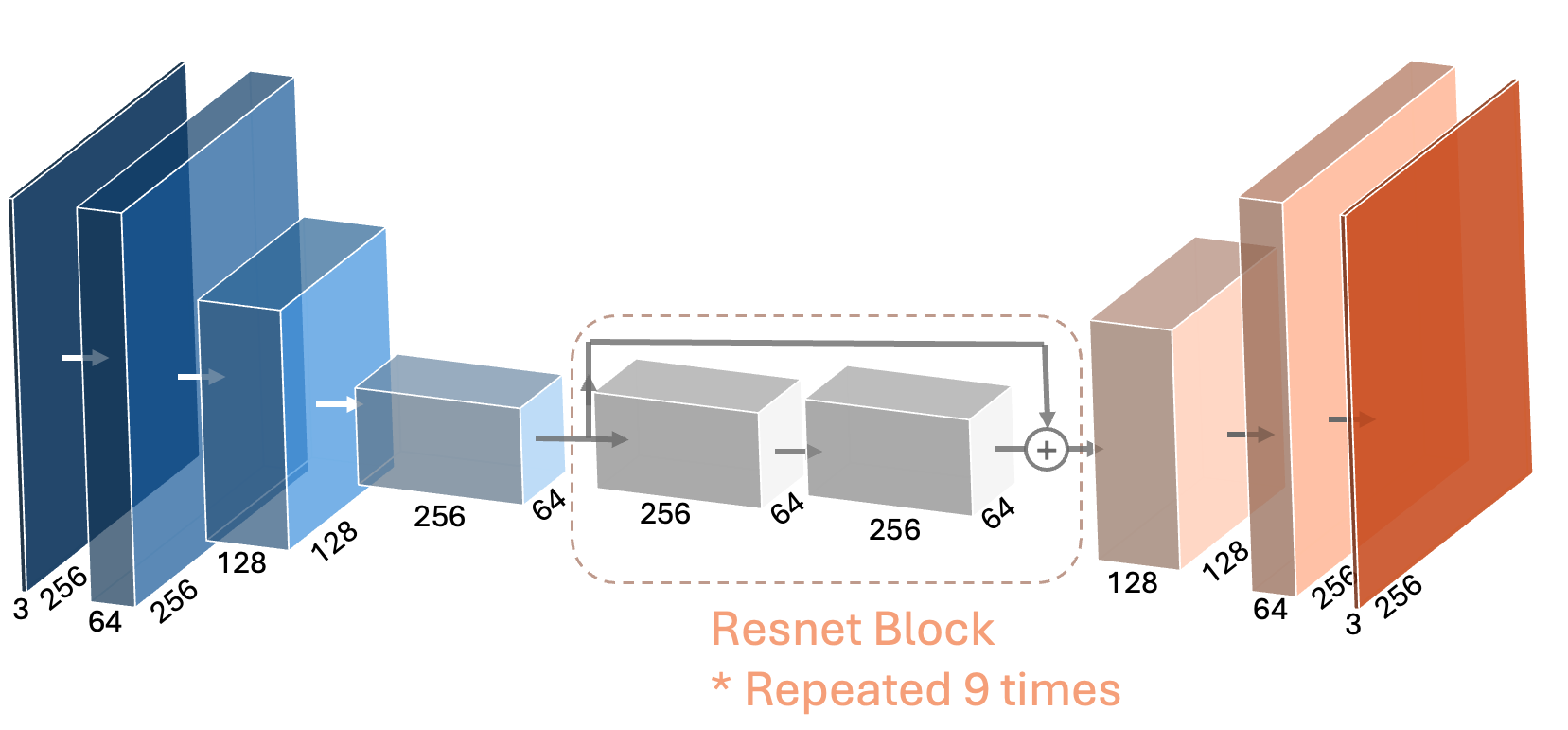}
    \caption{Architecture of the Cycle-GAN generator. The input is a 3-channel image of size $256\times256$, processed through convolutional layers. Between each layer, indicated by arrows, instance normalization and ReLU activations are applied. The input image is processed through the downsampling stage, followed by a sequence of residual (ResNet) blocks repeated 9 times. Then, the image is upsampled back to its original size, completing the transformation.}
    \label{fig: generator}
\end{figure}

\begin{figure}[!b]
    \centering
    \includegraphics[width=\linewidth]{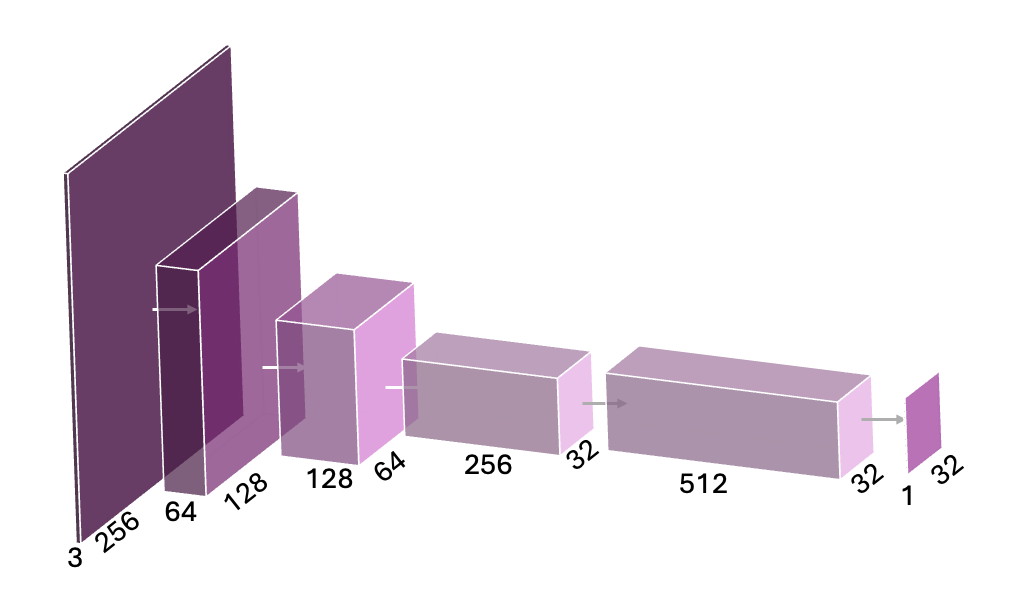}
    \caption{Architecture of the Cycle-GAN discriminator. The discriminator processes the input image through several downsampling layers, providing a $32\times32$ grid of decisions. In between layers, arrows indicate instance normalization and leaky ReLU.}
    \label{fig: discriminator}
\end{figure}

Cycle-GANs have two generator and discriminator networks as illustrated in Fig. \ref{fig:CGAN}. The first generator learns the forward mapping from the source domain $X$, consisting of echo recordings, to the target domain $Y$ representing the corresponding cardiac MRI sequences. Specifically, given sets of preprocessed echo $\{\mathbf{x}_i\}_{i=1}^{N_e}$ and MRI samples $\{\mathbf{y}_j\}_{j=1}^{N_m}$, where $\mathbf{x}_i \in X$, and $\mathbf{y}_j \in Y$, the generator mappings are defined as $G: X \rightarrow Y$ and $F: Y \rightarrow X$. We use the same network architectures for  $G$ and $F$ generators as depicted in Fig. \ref{fig: generator}. The discriminators corresponding to the generator outputs of $G$ and $F$ are expressed as $D_X$ and $D_Y$, respectively. These networks learn to distinguish the generated and real image samples from their corresponding target domain. The discriminator architecture used in this study is illustrated in Fig. \ref{fig: discriminator}.

Given an echocardiography frame $\mathbf{x} \in \mathbb{R}^{M \times N}$, each cycle starts with the following forward mapping: $\hat{\mathbf{y}} = G(\mathbf{x})$, where $\hat{\mathbf{y}} \in \mathbb{R}^{M \times N}$ is the generated cardiac MRI frame which is then fed to the discriminator, $D_Y$, to obtain a mask of $\hat{\mathbf{M}}_x = D_X(\hat{\mathbf{x}})$ where $\hat{\mathbf{M}}_x \in \mathbb{R}^{d_m \times d_n}$ expresses the pixel-wise decision of the discriminator as real or synthetic. While the generated images have the same size as the source domain samples, the mask dimensions are determined by the number of down-sampling operations in the discriminators. Accordingly, both generators have 18 convolutional layers equipped with residual connections, followed by the ReLu activation function and instance normalization. On the other hand, the discriminators have 5 convolutional layers followed by the leaky ReLu. The inverse transformation completes the one-cycle by obtaining the reconstructed given input echo frame $\tilde{\mathbf{x}} = F(\hat{\mathbf{y}})$. Then, the next cycle is computed with the generator $F$ where the source domain is now selected as $Y$, and an echocardiography frame is generated from the source cardiac MRI frame. The discriminator $D_X$ decides whether the generated frame is real or fake. These two cycles, $X \rightarrow Y \rightarrow X$ and $Y \rightarrow X \rightarrow Y$, form one iteration of the back-propagation during training.

The overall loss function of the Cycle-GAN model generator consists of three components, namely, two adversarial and one cycle-consistency loss which are expressed as $\mathcal{L}_A$, $\mathcal{L}_{\text{cyc}}$, respectively:
\begin{equation}
\label{eq:objective}
    \begin{split}
    \mathcal{L}_{\text{T}}(G, F, D_X, D_Y, & \mathbf{x}, \mathbf{y}) = \mathcal{L}_A(G, D_Y, \mathbf{x}, \mathbf{y}) \\
    & + \mathcal{L}_A(F, D_X, \mathbf{y}, \mathbf{x}) \\
    & + \lambda \mathcal{L}_{\text{cyc}}(G, F, \mathbf{x}, \mathbf{y}) \\
    \end{split}
\end{equation}

Generator and discriminator networks are trained in such a way that they compete with each other, for example, the weights of $G$ and $D_X$ are updated by $\min_G \mathcal{L}_A(G, D_Y, \mathbf{x}, \mathbf{y})$ and $\max_{D_X} \mathcal{L}_A(F, D_X, \mathbf{y}, \mathbf{x})$ objectives, respectively. Such adversarial training enables generating realistic MRI images that appear similar to the real MRI samples making it a challenging task for the discriminator to separate the synthesized images from the target domain samples (MRI views).

For a given set of $\mathbf{x} \in X$ and $\mathbf{y} \in Y$ samples, the generators separately calculate the mappings for their corresponding target domain images $\hat{{\mathbf{y}}} = G(\mathbf{x})$ and $\hat{{\mathbf{x}}} = F(\mathbf{y})$, then the following combined adversarial loss $\mathcal{L}_{A}^* $ in \eqref{eq:objective} is computed as follows:
\begin{equation}
\label{eq:gen-cost1}
    \begin{split}
    \mathcal{L}_{A}^*(D_Y, D_X, \hat{\mathbf{y}}, \hat{\mathbf{x}}) = & \mathcal{L}_A(G, D_Y, \mathbf{x}, \mathbf{y}) + \mathcal{L}_A(F, D_X, \mathbf{y}, \mathbf{x}) \\
    = &\left \| D_Y\left( \hat{\mathbf{y}} \right) - \mathbf{1} \right \|_2^2 + \left \| D_X\left( \hat{\mathbf{x}} \right) - \mathbf{1} \right \|_2^2.
    \end{split}
\end{equation}
Next, the inverse transformations are applied to the generated images: $\tilde{\mathbf{x}} = F(\hat{\mathbf{y}})$ and $\tilde{\mathbf{y}} = G(\hat{\mathbf{x}})$, and the cycle-consistency loss is expressed as,
\begin{equation}
\label{eq:cyc-cost}
    \mathcal{L}_{\text{cyc}}(\mathbf{x}, \tilde{\mathbf{x}}, \mathbf{y}, \tilde{\mathbf{y}}) = \left \| \tilde{\mathbf{x}} - \mathbf{x} \right \|_1 + \left \| \tilde{\mathbf{y}} - \mathbf{y} \right \|_1.
\end{equation}

Given $G$, $F$, and the sample set of $\{\mathbf{x}, \mathbf{y}\}$, the discriminator losses are expressed for $D_X$ and $D_Y$ as follows,
\begin{equation}
\label{eq:dis1}
    \mathcal{L}_{D_1}(G, D_Y, \mathbf{x}, \mathbf{y}) = \left\| D_Y(\mathbf{y}) - \mathbf{1} \right\|_2^2 + \left\| D_Y(G(\mathbf{x}))\right\|_2^2,
\end{equation}
\begin{equation}
\label{eq:dis2}
    \mathcal{L}_{D_2}(F, D_X, \mathbf{y}, \mathbf{x}) = \left\| D_X(\mathbf{x}) - \mathbf{1} \right\|_2^2 + \left\| D_X(F(\mathbf{y}))\right\|_2^2,
\end{equation}
and the total loss $L_D = L_{D_1} + L_{D_2}$. Consequently, in the training of the Cylce-GANs, the minimization of $L_D$ is equivalent to maximizing $L_A$ \c{eq:gen-cost1} for the discriminators.

\section{Experimental Results}
\label{sec: experimental}

This section first introduces the Echo2MRI dataset, encapsulating paired echocardiography and cardiac MRI data. Then, we will outline the experimental setup used to evaluate the proposed method. Following this, we will present both qualitative assessment as well as and three sets of quality evaluations performed independently by a group of expert cardiologists. These expert evaluations shall demonstrate that the proposed MRI transformation can maintain clinical relevance, reliability, and accuracy, ensuring its potential for real cardiac diagnostics. Lastly, we will present the computational complexity analysis of the proposed method.

\subsection{Echo2MRI Dataset}

The Echo2MRI dataset was created in collaboration with Hayatabad Medical Complex in Pakistan. The data was collected anonymously with the approval of the local ethics board of Hayatabad Medical Complex, which granted authorization in February 2019.
The dataset comprises apical 4-chamber views of paired echocardiography and cardiac MRI views from 36 patients. In addition, 4 patients have only cardiac MRI data, and 12 patients have only echocardiography data.

\begin{figure}[t]
    \centering
    \includegraphics[width=\linewidth]{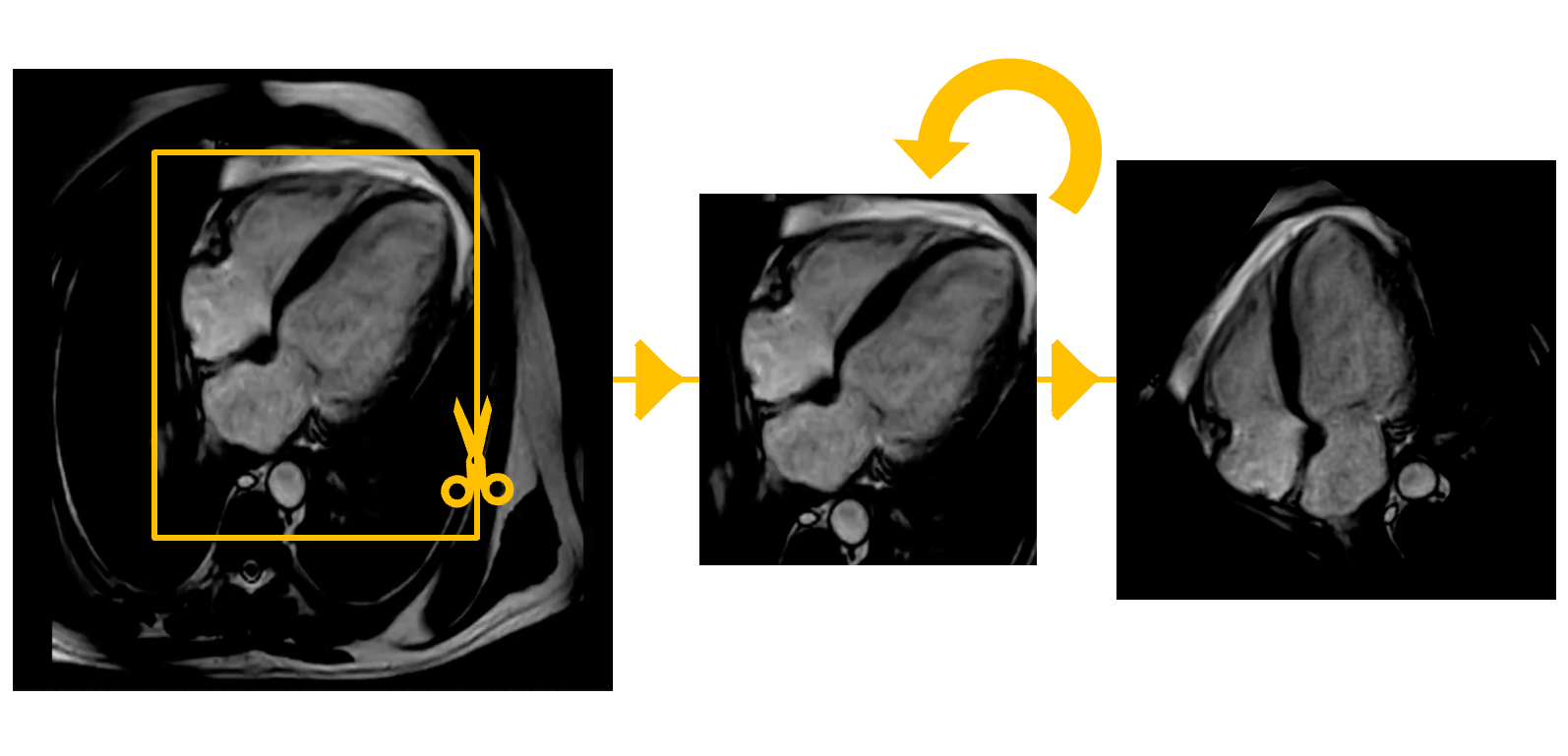}
    \caption{Initial cardiac MRIs are cropped to focus on the apical 4-chamber cardiac view and neighboring veins, then rotated to position the left ventricle in the upper left corner, aligning with the echocardiography view.}
    \label{fig:cutrotate}
\end{figure}

To compose the Echo2MRI dataset, we first extracted the regions containing the apical 4-chamber view from both echocardiogram and cardiac MRI videos. To simplify the learning, we aligned the cardiac MRI videos by rotating the cropped regions to roughly match the orientation of the echocardiograms, ensuring the left ventricle (LV) is positioned towards the top-right corner. Then, we extracted the frames containing 4-chamber cardiac views from these videos.
Fig. \ref{fig:cutrotate} illustrates the pre-processing steps over the cardiac MRI frames.

To ensure the best transformation quality in the cardiac MRI domain, which will naturally provide the best echocardiography restoration, we conducted a visual assessment. We selected the 16 best-quality cardiac MRI videos for the training process. Among these 16 best-quality videos, 12 have paired echocardiography and 4 have only MRI.
Pairs of selected MRI videos and an additional 12 unpaired echocardiography videos are used for training.
 Overall, the training data covers echocardiography from 24 patients and cardiac MRI from 16 patients.
From these videos, training data includes 1138 high-quality cardiac MRI frames and 1193 echocardiography frames, with additional data augmentations, such as rotation.
Furthermore, we randomly flipped half of the frames to enhance diversity and achieve a more robust model.
 The rest of the paired data, 24 patients, are spared for testing. Cardiac MRI and echocardiography videos have different frame rates; therefore, we spatially registered and matched the temporal movement across echocardiography and cardiac MRI frames. 
 In total, we used 855 echocardiography frames and 855 corresponding cardiac MRI frames for testing.
The data preparation and evaluation format will be detailed in the following subsections.
 
\subsection{Experimental Setup}

We used a batch size of 1 and a learning rate 0.0002 to train the Cycle-GAN model. The model was trained for 200 epochs, with a linearly decaying learning rate to zero starting at epoch 100. We employed Adam optimizer with $\beta_1=0.5$. The cycle-consistency loss weight, $\lambda$ in \eqref{eq:objective}, was set to 10. 
We used Python with the TensorFlow library \cite{abadi2016tensorflow}.

\subsection{Qualitative Evaluation}

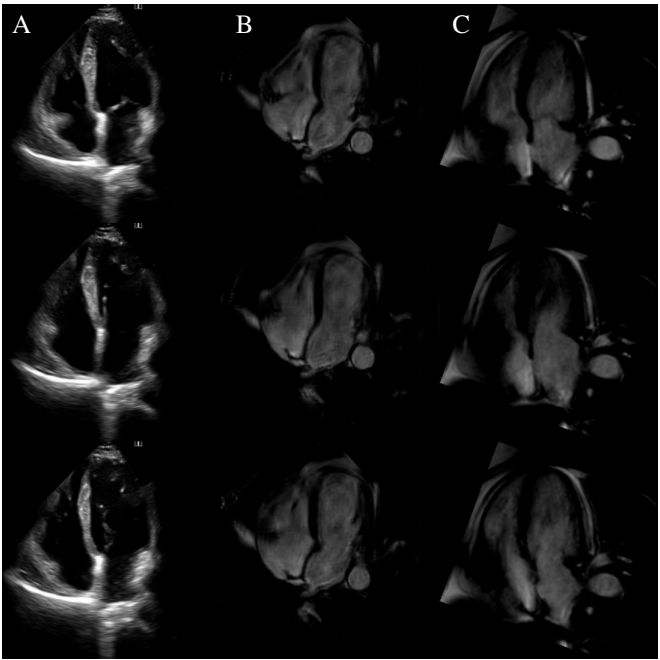
\begin{figure}[!t]
    \centering
    \begin{tikzpicture}
        \node[anchor=south west, inner sep=0] (image) at (0,0) {\includegraphics[width=0.48\textwidth]{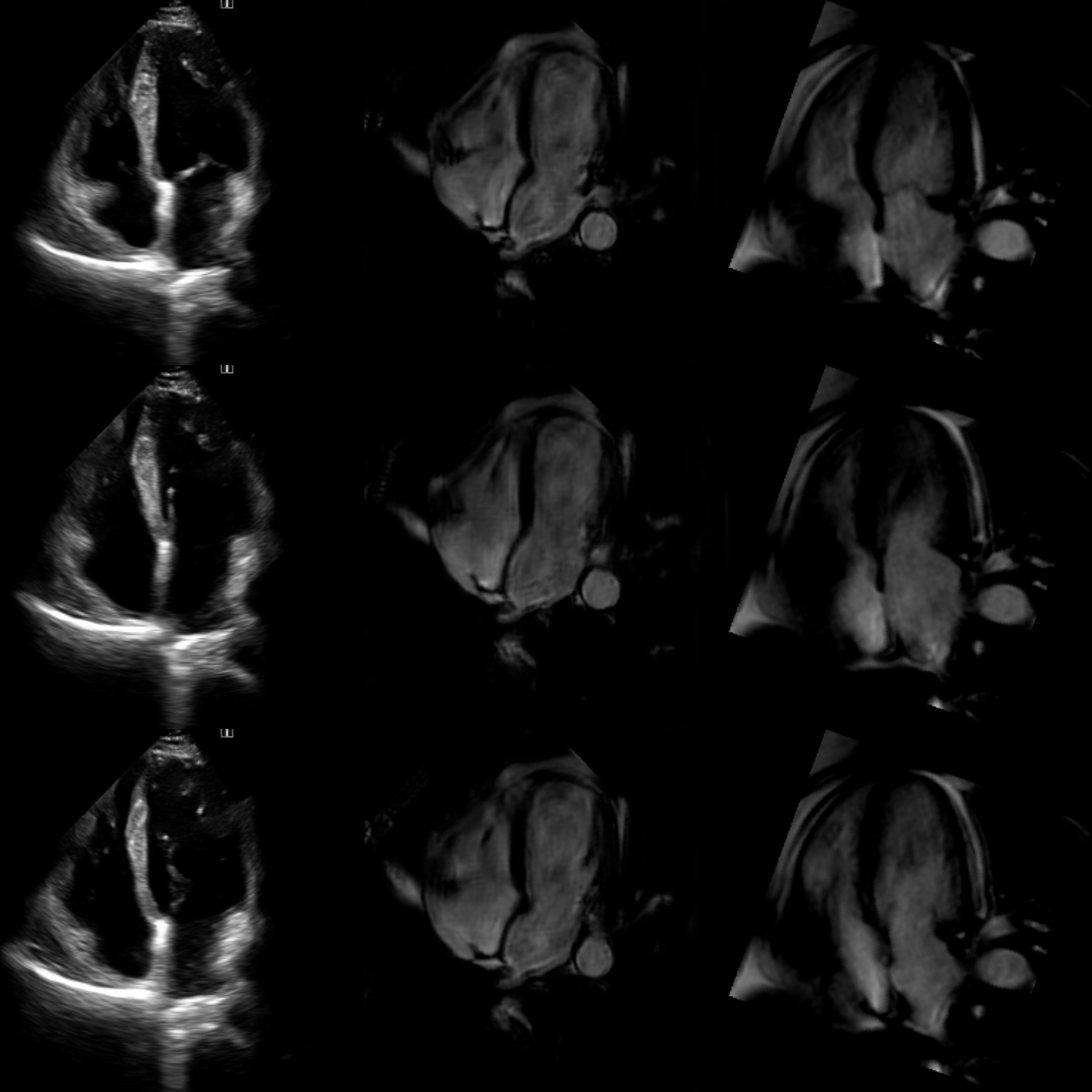}}; 
        \begin{scope}[x={(image.south east)}, y={(image.north west)}]
            \node[anchor=center, text=white] at (0.03, 0.97) {A};
            \node[anchor=center, text=white] at (0.37, 0.97)  {B};
            \node[anchor=center, text=white] at (0.70, 0.97) {C};
        \end{scope}
    \end{tikzpicture}
    \caption{Echocardiography (column A), generated cardiac MRI (column B), and original cardiac MRI (column C) frames across different points of the cardiac cycle for $10^{th}$ patient.}
    \label{Fig: qualitative}
\end{figure}

To assess the visual quality of the generated cardiac MRI views, we conducted a qualitative evaluation based on several key criteria. Using the test set, we performed echocardiography frame transformation, and compared it with its original echocardiography and time-registered cardiac MRI views as shown in Fig. \ref{Fig: qualitative}. 

First, we check whether the generated views exhibit clear boundaries delineating the cardiac walls. Well-defined boundaries are crucial for the accurate assessment of RWMA and cardiac anatomical features by both cardiologists and machines. Especially when certain LV wall segments are out of view in the source domain echocardiography, the transformed cardiac MRI view should be able to synthesize the complete LV wall in its structural integrity.

Secondly, we check whether the generated views represent the actual cardiac motion throughout the cardiac cycle. This includes ensuring the smooth and consistent transition between systole and diastolic phases and the corresponding movement of the muscles. The consistent transformation of muscle movement's pace indicates the models can incorporate dynamic information through temporal space. This is especially crucial for RWMA detection to diagnose any onset of myocardial ischemia and infarction (MI).

Finally, we consider the overall visual consistency and quality of the generated cardiac MRI views. This involved checking for artifacts, smoothness, and consistency of the visible arteries and other cardiac features. This is particularly important to validate that the model is not hallucinating while generating the view, but incorporating the information from the source view. 

Some typical transformation results over the test set are shown in Fig. \ref{Fig: qualitative} along with the source and target views. An extensive set of additional results from the test set are provided in Appendix A\footnote{The rest of the results are available in Appendix A.} and also in video format online\footnote{\href{https://github.com/ilkeadalioglu/Echo2MRI}{https://github.com/ilkeadalioglu/Echo2MRI}}.

Results indicate that the quality of the transformed cardiac MRI views can mostly satisfy the aforementioned criteria and usually present a similar or sometimes even better visual quality than the corresponding actual (target) cardiac MRI views. To validate this claim, we further performed a ``Confusion Test" as part of the domain expert evaluation, which will be presented next. 

\subsection{Domain Expert Evaluations}

Four cardiologists perform each of the three evaluations independently, and the evaluation scores are averaged and reported in this section. The first two tests reveal crucial expert assessments of the quality and importance of the proposed echocardiography restoration by cardiac MRI view transformation. The last test demonstrates a one-to-one comparison between the transformed and actual cardiac MRI time sequences. The rest of the section details each evaluation and presents their results with important remarks.

\subsubsection{Confusion test}
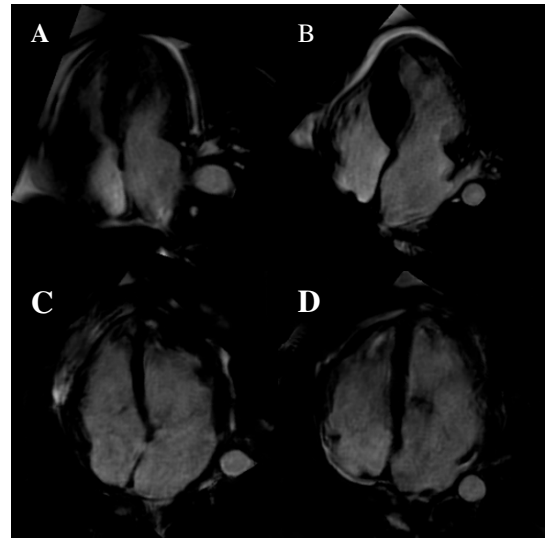
\begin{figure}[t]
    \centering
    \begin{tikzpicture}
        \node[anchor=south west, inner sep=0] (image) at (0,0) {
            \includegraphics[width=0.8\linewidth]{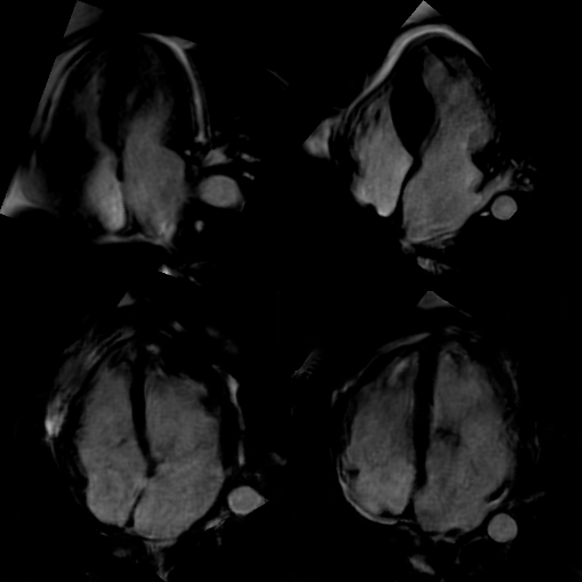}
        };
        \begin{scope}[x={(image.south east)}, y={(image.north west)}]
            \node[anchor=north west, text=white, font=\bfseries] at (0.02, 0.98) {A};
            \node[anchor=north west, text=white] at (0.52, 0.98) {B};
            \node[anchor=north west, text=white, font=\large\bfseries] at (0.02, 0.48) {C};
            \node[anchor=north west, text=white, font=\large\bfseries] at (0.52, 0.48) {D};
        \end{scope}
    \end{tikzpicture}
    \caption{A sample from the confusion test containing four randomly selected, generated, or original MRI images. Cardiologists were asked to identify the generated ones.\protect\footnotemark}
    \label{fig:exp1_example}
\end{figure}
\footnotetext{Labels: A\&B, original; C\&D, synthesized.}

The first question we asked cardiologists was to identify which of the shown views are synthetic. We call this first evaluation test the ''Confusion Test". As its name implies, we aimed to assess how well a trained expert's eye can distinguish the transformed (synthetic) cardiac MRI frame from the original counterpart.  Each cardiologist were shown  200 cardiac MRI views, consisting of 100 synthetic and 100 original cardiac MRI images, and were tasked with identifying the synthetic ones. Frames are shown to the expert randomly, ensuring that each frame has a probability of 0,5 to be an original or generated view. Therefore, if the quality of the transformed cardiac MRI views matches their actual counterparts, one can expect an average accuracy of $0.5$ or lower, otherwise, the accuracy can be expected to be higher than $0.5$. Other standard evaluation metrics such as Precision, Recall, Specificity, F1 score, false positive rate (FPR), and Matthews correlation coefficient (MCC) are also computed for an overall quantitative evaluation.  Figure \ref{fig:exp1_example} shows a sample from the Confusion Test\footnote{More examples are available in Appendix B.}. We retrieved each assessment from the four cardiologists and presented results in a confusion matrix (CM). 

\begin{table}[h]
\caption{Confusion matrix of the evaluation test 1: Confusion Test. Classification results of 4 cardiologists over randomly selected 200 test images, 100 original and 100 synthetic cardiac MRI views, with $1$ indicating synthetic (positive), and $0$ indicating original (negative).}
\renewcommand{\arraystretch}{1.2} 
\setlength{\tabcolsep}{12pt} 
\resizebox{0.8\columnwidth}{!}{%
\begin{tabular}{cccc}
&  &  \multicolumn{2}{c}{\textbf{Actual Values}}  \\ \cline{3-4} 
& \multicolumn{1}{l|}{} & \multicolumn{1}{c|}{\cellcolor[HTML]{FE0000}\textbf{1}} & \multicolumn{1}{c|}{\textbf{0}}   \\ \cline{2-4} 
\multicolumn{1}{l|}{\textbf{Predicted Values}} & \multicolumn{1}{l|}{\cellcolor[HTML]{FE0000}\textbf{1}} & \multicolumn{1}{c|}{\textbf{68}} & \multicolumn{1}{c|}{\textbf{129}} \\ \cline{2-4} 
\multicolumn{1}{l|}{} & \multicolumn{1}{l|}{\textbf{0}}  & \multicolumn{1}{c|}{\textbf{332}} & \multicolumn{1}{c|}{\textbf{271}} \\ \cline{2-4} 
\end{tabular}}
\label{Tab:CM}
\end{table}

True positives, (TP) are correctly classified synthetics; false positives (FP) are originals, incorrectly classified as synthetic; false negatives (FN) are synthetics, incorrectly classified as original; and finally true negatives (TN) are correctly classified originals.
From the CM, we calculate \textit{accuracy}, the ratio of correctly identified samples; \textit{precision}, the rate of correctly identified synthetics amongst all classified as synthetic; \textit{recall}, the proportion of correctly detected synthetics, amongst all synthetics; \textit{false positive rate} (FPR) is the rate of false alarm in classifying synthetics. \textit{F1} score represents the harmonic mean of precision and recall as in Eq. \ref{eqn:F1}; and finally Matthews Correlation Coefficient (MCC) is the quality of the dependence, as expressed in Eq. \ref{eqn:MCC}.

\begin{equation}
F1 = 2 \times \frac{\text{Precision} \times \text{Recall}}{\text{Precision} + \text{Recall}}
\label{eqn:F1}
\end{equation}

\begin{equation}
\text{MCC} = \frac{(TP \times TN) - (FP \times FN)}{\sqrt{(TP + FP)(TP + FN)(TN + FP)(TN + FN)}}
\label{eqn:MCC}
\end{equation}

\begin{table}[h]
\caption{Synthetic detection performances of 4 cardiologists over randomly selected 200 test images, 100 original and 100 synthetic cardiac MRI.}
\renewcommand{\arraystretch}{1.2} 
\resizebox{\columnwidth}{!}{%
\Large
\begin{tabular}{c|c|c|c|c|c|c}
\textbf{Accuracy} & \textbf{Precision} & \textbf{Recall} & \textbf{Specificity} & \textbf{F1 Score} & \textbf{MCC} & \textbf{FPR} \\ \hline
\textbf{0.47} & \textbf{0.45} & \textbf{0.33} & \textbf{0.60} & \textbf{0.38} & \textbf{-0.07} & \textbf{0.40} 
\end{tabular}}
\label{Tab:metrics}
\end{table}

The evaluation results presented in Table II indicate that cardiologists could not accurately discriminate between synthetic and real cardiac MRI images. In particular, the accuracy of $0.47$ and MCC of $-0.007$ demonstrate that the cardiologists' performance in identifying synthetic images is almost random, as accuracy close to $0.5$ is as trustworthy as flipping a coin and MCC close to $0$ shows no correlation in their identification. A precision of 0.45 suggests that only less than half of the synthetic images were identified as actually synthetic. A recall of 0.33 reveals that only one-third of the actual synthetic cardiac MRIs were correctly identified as synthetic, and cardiologists frequently confused synthetic images as real. Overall, such high confusion shows that the proposed approach can synthesize such synthetic cardiac MRI frames that are quite similar to the actual ones.

\subsubsection{RWMA Test} 

The second evaluation test is the "Regional Wall Motion Abnormality (RWMA) Test". This test aims to determine whether the transformed cardiac MRI views can capture the important temporal and structural information of the echocardiography and in particular, the LV wall motion for detecting the RWMA. Cardiologists were presented with echocardiography and corresponding synthetic cardiac MRI videos and asked to indicate which one they would prefer for RWMA diagnosis: echocardiography, synthetic cardiac MRI, or both; meaning they carry the same or similar diagnostic information. If they chose echocardiography, they were also asked if they would prefer to see its synthetic cardiac MRI alongside, for additional information.

We collected 96 evaluations from the four cardiologists over 24 test patients. One evaluation was omitted because the cardiologist omitted both views due to their low diagnostic quality.

\begin{table}[h]
\caption{Cardiologists' preferences when asked which view is preferred to base their diagnosis on; echocardiography, synthetic cardiac MRI, or both views provide similar information. The table represents the percentage of responses for each option.}
\renewcommand{\arraystretch}{1.2} 
\resizebox{\columnwidth}{!}{%
\Large
\begin{tabular}{c|c|c}
\multirow{2}{*}{\textbf{Echocardiography}} & \multirow{2}{*}{\textbf{Synthetic cardiac MRI}} & \multirow{2}{*}{\begin{tabular}[c]{@{}c@{}}\textbf{Both Echocardiography}\\ \textbf{\& Synthetic cardiac MRI}\end{tabular}} \\
 & & \\ \hline
\textbf{35.8\%}  & \textbf{29.5\%} & \textbf{34.7\% }
\end{tabular}}
\label{Tab:exp2}
\end{table}

Results in Table \ref{Tab:exp2} reveal that the transformed cardiac MRI views can present a similar or better RWMA analysis on the majority of the cases (\textbf{64.2\%}). Only in \textbf{35.8\%} cases, echocardiography was found superior to synthetic cardiac MRI in diagnostic quality. However, when asked if the synthetic cardiac MRI should be seen alongside the echocardiography for additional information, cardiologists responded positively in \textbf{41.2\%} of the cases. This suggests that while cardiologists may rely on echocardiography for their primary diagnosis, a significant portion also wanted to see synthetic cardiac MRI alongside for additional information. 

Cardiologists noted that three echocardiography videos exhibit missing movements, which were also unclear in their corresponding synthetic cardiac MRI translations. In particular, missing key features such as the septal wall, or displaying dyskinetic movements hinder accurate diagnosis. It is important to recognize that very low-quality echocardiography can still be challenging to restore. In such cases, synthetic cardiac MRI may carry the same diagnostic limitations as the source echocardiography. 
However, it is also important to note that, echocardiography can be successfully restored through the transformation to synthetic Cardiac MRI, resulting in improved diagnostic quality, as in \textbf{29.5\%} of the cases.   

Overall, synthetic cardiac MRI is assessed to be effective and provide additional information in \textbf{78.9\%} of the cases. This highlights the significance synthetic cardiac MRI may bring to the diagnostic process and demonstrates its potential to enhance accuracy in cardiac diagnosis, especially in detecting RWMA.

\subsubsection{Quality Test}
The third evaluation test is called the ``Quality Test", which aims to evaluate the video quality and relevance of synthetic cardiac MRI compared to the actual cardiac MRI. In other words, this test aims to assess the model's ability to accurately learn the transformation domain to generate synthetic cardiac MRIs of comparable quality. In this test, cardiologists were presented with synthetic and original cardiac MRI videos and asked to evaluate the quality of the synthetic cardiac MRI compared to the original by selecting one of the following answers: Similar, Better, or Worse.  

\begin{table}[h]
\centering
\caption{Cardiologists' assessments of the quality of synthetic cardiac MRI compared to original cardiac MRI. The table shows the percentage of the responses where the synthetic cardiac MRI was evaluated as better, similar, or worse than the original.}
\renewcommand{\arraystretch}{1.2} 
\resizebox{0.7\columnwidth}{!}{%
\scriptsize
\begin{tabular}{c|c|c}
\textbf{Better} & \textbf{Similar} & \textbf{Worse} \\ \hline
\textbf{14.58\% } & \textbf{30.21\%} & \textbf{55.21\%} 
\end{tabular}}
\label{Tab:exp3}
\end{table}

Table \ref{Tab:exp3} indicates that cardiologists rated synthetic cardiac MRIs better than the original in \textbf{14.58\%} of the cases. In this case, the model was able to transform the input echocardiography to its cardiac MRI domain representation, with diagnostic quality and clarity exceeding that of the original cardiac MRI. Also, in \textbf{30.21\%} of the cases, synthetic cardiac MRI was found to be of similar quality to the original cardiac MRI. Overall, in \textbf{44.79\%} of the total cases, synthetic cardiac MRI were found to be equal to or better than the original cardiac MRI. On the other hand, \textbf{55.21\%} of the cases were rated worse than the original cardiac MRI.

Creating superior cardiac MRI views than their original counterparts was not the main objective of this study. 
Due to the discordant movements between echocardiography and MRI \cite{uretsky2015discordance_dyskinetic}, the synthetic counterpart also exhibits a lack of harmony after translation. The difference in the movements highlights the room for improvement. Further refinement is needed to address the quality gap that causes dyskinetic movements occurring in low-quality echocardiography inputs.

\subsection{Computational Complexity Analysis}
The generator $G:X\rightarrow Y$ that translates echocardiography to cardiac MRI consists of approximately $11.383$ million trainable parameters. The model's inference was evaluated under two different setups. On Intel(R) Xeon(R) Gold 6230 CPU @ 2.10GHz, the average inference time per frame is $1.2886$ seconds. However, a Tesla V100-SXM2-32GB GPU significantly reduces the average inference time to $0.0184$ seconds. This setup enables real-time translation, exceeding the original frame rate of echocardiography at 58.82, hence, achieving cardiac MRI domain videos, with a frame rate even higher than the original cardiac MRI.

\section{Conclusion}
\label{sec: conclusion}
This pilot study proposes a transformation-based restoration approach to restore echocardiography by transforming into a synthetic cardiac MRI view, as MRIs are considered the gold standard in cardiac imaging. The proposed method blindly restores the noisy echo without any prior knowledge about the type or severity of the corrupting artifacts. To accomplish this aim, we created the first multi-view public dataset of echocardiography and cardiac MRI with carefully designed pre- and post-processing steps.

Although RWMA is the first abnormality to set in with the onset of myocardial ischemia, preceding metabolic and electrocardiographic abnormalities, currently it is usually used as the secondary diagnostic tool in patients with non-diagnostic ECG or when the diagnosis is not evident by “standard” means. This is despite the fact that echocardiography, particularly myocardial strain imaging, provides a highly sensitive and specific, earliest diagnosis of MI by detecting an ongoing RWMA. A major reason for not using echocardiography as a first action line diagnostic tool, for suspected MI patients is the blend of artifacts which often makes it infeasible to make a reliable diagnosis by both cardiologists and prior machine learning paradigms proposed to date. 

Three evaluation tests demonstrate that the transformed cardiac MRI views effectively suppressed the artifacts present in the source echocardiography in general. Even when the echocardiography has some obscured LV wall segments, the transformed cardiac MRI view was able to generate the entire LV wall, which is crucial for RWMA analysis. In brief, the delineation of the cardiac movement and anatomy is satisfactory and consistent, providing sufficient quality for an accurate diagnosis of cardiac disorders. In particular, visualizing LV wall accurately helps cardiologists and opens up space for machine learning research to improve early MI detection using echocardiography \cite{kiranyaz_LV}.

The confusion test revealed that the transformed cardiac MRI view is indistinguishable from the actual counterpart.  RWMA test further shows that in \textbf{78.9\%} of the cases, cardiologists prefer the transformed MRI view for RWMA detection. Finally, in the quality test, the transformed cardiac MRI quality can be equal to or even higher than the actual counterpart in \textbf{44.79\%} of the cases. 

Despite the model's success in restoring corrupted echocardiography via transformation, it is important to note that high levels of corruption in echocardiography can still hinder the generation of realistic cardiac patterns and motion in synthetic cardiac MRI images. Future studies will focus on enhancing the model's ability to maintain temporal coherency, further improving synthetic cardiac MRI images' diagnostic value and reliability to converge to their actual counterparts.

\section*{Acknowledgment}
We want to thank Hayatabad Medical Complex for their contribution to collecting the dataset. Additionally, we extend our gratitude to Majid Khan, Hayatabad Medical Complex, Pakistan; Dr. Kamran Majeed (MBBS, PhD, FRACP, FCSANZ), Waikato Hospital, New Zealand; Dr. Sabir Abdul Karim (MBBS, MRCP, FRCP[Edin], FEACVI, FSCMR), Rotherham General Hospital, UK; Dr. Awad Al-Qahtani, Hamad Medical Corporation Heart Hospital, Qatar, for their valuable contribution to the qualitative evaluation.

We want to acknowledge CSC; IT Center for Science, Finland, for the computational resources.

\bibliographystyle{IEEEtran}
\bibliography{main}

\begin{thebibliography}{10}
\providecommand{\url}[1]{#1}
\csname url@samestyle\endcsname
\providecommand{\newblock}{\relax}
\providecommand{\bibinfo}[2]{#2}
\providecommand{\BIBentrySTDinterwordspacing}{\spaceskip=0pt\relax}
\providecommand{\BIBentryALTinterwordstretchfactor}{4}
\providecommand{\BIBentryALTinterwordspacing}{\spaceskip=\fontdimen2\font plus
\BIBentryALTinterwordstretchfactor\fontdimen3\font minus \fontdimen4\font\relax}
\providecommand{\BIBforeignlanguage}[2]{{%
\expandafter\ifx\csname l@#1\endcsname\relax
\typeout{** WARNING: IEEEtran.bst: No hyphenation pattern has been}%
\typeout{** loaded for the language `#1'. Using the pattern for}%
\typeout{** the default language instead.}%
\else
\language=\csname l@#1\endcsname
\fi
#2}}
\providecommand{\BIBdecl}{\relax}
\BIBdecl

\bibitem{WHO}
\BIBentryALTinterwordspacing
W.~H. Organization, Jun 2021. [Online]. Available: \url{https://www.who.int/news-room/fact-sheets/detail/cardiovascular-diseases-(cvds)}
\BIBentrySTDinterwordspacing

\bibitem{counseller2023recent}
Q.~Counseller and Y.~Aboelkassem, ``Recent technologies in cardiac imaging,'' \emph{Frontiers in Medical Technology}, vol.~4, p. 984492, 2023.

\bibitem{dalen2014CAD}
J.~E. Dalen, J.~S. Alpert, R.~J. Goldberg, and R.~S. Weinstein, ``The epidemic of the 20th century: coronary heart disease,'' \emph{The American journal of medicine}, vol. 127, no.~9, pp. 807--812, 2014.

\bibitem{pennell2004_5_5xcost}
D.~J. Pennell, U.~P. Sechtem, C.~B. Higgins, W.~J. Manning, G.~M. Pohost, F.~E. Rademakers, A.~C. van Rossum, L.~J. Shaw, and E.~K. Yucel, ``Clinical indications for cardiovascular magnetic resonance (cmr): Consensus panel report,'' \emph{European heart journal}, vol.~25, no.~21, pp. 1940--1965, 2004.

\bibitem{bullock2024multimodality}
R.~P. Bullock-Palmer, K.~Flores~Rosario, P.~S. Douglas, R.~T. Hahn, R.~M. Lang, P.~Chareonthaitawee, M.~B. Srichai, K.~G. Ordovas, L.~A. Baldassarre, M.~S. Burroughs \emph{et~al.}, ``Multimodality cardiac imaging and the imaging workforce in the united states: diversity, disparities, and future directions,'' \emph{Circulation: Cardiovascular Imaging}, vol.~17, no.~2, p. e016409, 2024.

\bibitem{kurtz1981echogenicity}
A.~Kurtz, P.~Dubbins, C.~Rubin, R.~Kurtz, H.~Cooper, C.~Cole-Beuglet, and B.~Goldberg, ``Echogenicity: analysis, significance, and masking,'' \emph{American Journal of Roentgenology}, vol. 137, no.~3, pp. 471--476, 1981.

\bibitem{wu2015echocardiogram|_multiple_sources_of_noise_data_driven_denoising}
H.~Wu, T.~T. Huynh, and R.~Souvenir, ``Echocardiogram enhancement using supervised manifold denoising,'' \emph{Medical image analysis}, vol.~24, no.~1, pp. 41--51, 2015.

\bibitem{kinno2017comparison}
M.~Kinno, P.~Nagpal, S.~Horgan, and A.~H. Waller, ``Comparison of echocardiography, cardiac magnetic resonance, and computed tomographic imaging for the evaluation of left ventricular myocardial function: part 2 (diastolic and regional assessment),'' \emph{Current cardiology reports}, vol.~19, pp. 1--13, 2017.

\bibitem{votavova2015LVecho}
R.~Votavov{\'a}, A.~Linhartov{\'a}, J.~Ko{\v{r}}{\'\i}nek, J.~Marek, and A.~Linhart, ``Echocardiography in coronary artery disease,'' \emph{Cor et Vasa}, vol.~57, no.~6, pp. e408--e418, 2015.

\bibitem{diller2019denoising}
G.-P. Diller, A.~E. Lammers, S.~Babu-Narayan, W.~Li, R.~M. Radke, H.~Baumgartner, M.~A. Gatzoulis, and S.~Orwat, ``Denoising and artefact removal for transthoracic echocardiographic imaging in congenital heart disease: utility of diagnosis specific deep learning algorithms,'' \emph{The international journal of cardiovascular imaging}, vol.~35, pp. 2189--2196, 2019.

\bibitem{khare2010wavelet_denoising}
A.~Khare, M.~Khare, Y.~Jeong, H.~Kim, and M.~Jeon, ``Despeckling of medical ultrasound images using daubechies complex wavelet transform,'' \emph{Signal Processing}, vol.~90, no.~2, pp. 428--439, 2010.

\bibitem{gupta2004wavelet}
S.~Gupta, R.~Chauhan, and S.~Sexana, ``Wavelet-based statistical approach for speckle reduction in medical ultrasound images,'' \emph{Medical and Biological Engineering and computing}, vol.~42, pp. 189--192, 2004.

\bibitem{collins2015globalCMRgold}
J.~D. Collins, ``Global and regional functional assessment of ischemic heart disease with cardiac mr imaging,'' \emph{Radiologic Clinics}, vol.~53, no.~2, pp. 369--395, 2015.

\bibitem{salerno2017gold_standards}
M.~Salerno, B.~Sharif, H.~Arheden, A.~Kumar, L.~Axel, D.~Li, and S.~Neubauer, ``Recent advances in cardiovascular magnetic resonance: techniques and applications,'' \emph{Circulation: Cardiovascular Imaging}, vol.~10, no.~6, p. e003951, 2017.

\bibitem{bruder2005detection}
O.~Bruder, K.~Waltering, P.~Hunold, M.~Jochims, B.~Narin, G.~Sabin, and J.~Barkhausen, ``Detection and characterization of left ventricular thrombi by mri compared to transthoracic echocardiography,'' \emph{RoFo: Fortschritte auf dem Gebiete der Rontgenstrahlen und der Nuklearmedizin}, vol. 177, no.~3, pp. 344--349, 2005.

\bibitem{murali2023bringing34}
S.~Murali, H.~Ding, F.~Adedeji, C.~Qin, J.~Obungoloch, I.~Asllani, U.~Anazodo, N.~A. Ntusi, R.~Mammen, T.~Niendorf \emph{et~al.}, ``Bringing mri to low-and middle-income countries: directions, challenges and potential solutions,'' \emph{NMR in Biomedicine}, p. e4992, 2023.

\bibitem{zhu2017unpairedCycleGan}
J.-Y. Zhu, T.~Park, P.~Isola, and A.~A. Efros, ``Unpaired image-to-image translation using cycle-consistent adversarial networks,'' in \emph{Proceedings of the IEEE international conference on computer vision}, 2017, pp. 2223--2232.

\bibitem{abadi2016tensorflow}
M.~Abadi, P.~Barham, J.~Chen, Z.~Chen, A.~Davis, J.~Dean, M.~Devin, S.~Ghemawat, G.~Irving, M.~Isard \emph{et~al.}, ``$\{$TensorFlow$\}$: a system for $\{$Large-Scale$\}$ machine learning,'' in \emph{12th USENIX symposium on operating systems design and implementation (OSDI 16)}, 2016, pp. 265--283.

\bibitem{uretsky2015discordance_dyskinetic}
S.~Uretsky, L.~Gillam, R.~Lang, F.~A. Chaudhry, E.~Argulian, A.~Supariwala, S.~Gurram, K.~Jain, M.~Subero, J.~J. Jang \emph{et~al.}, ``Discordance between echocardiography and mri in the assessment of mitral regurgitation severity: a prospective multicenter trial,'' \emph{Journal of the American College of Cardiology}, vol.~65, no.~11, pp. 1078--1088, 2015.

\bibitem{kiranyaz_LV}
S.~Kiranyaz, A.~Degerli, T.~Hamid, R.~Mazhar, R.~E. Fadil~Ahmed, R.~Abouhasera, M.~Zabihi, J.~Malik, R.~Hamila, and M.~Gabbouj, ``Left ventricular wall motion estimation by active polynomials for acute myocardial infarction detection,'' \emph{IEEE Access}, vol.~8, pp. 210\,301--210\,317, 2020.

\end{thebibliography}

\newpage
\onecolumn
\section*{\textbf{Appendix A}: Testing Samples for Echocardiography, Synthetic Cardiac MRI, and Real Cardiac MRI Comparison}
\label{sec: app1}
\begin{figure}[!h]
    \centering
    \begin{minipage}[t]{0.49\textwidth}
        \centering
        \begin{tikzpicture}
            \node[anchor=south west, inner sep=0] (image) at (0,0) {%
              \includegraphics[width=\linewidth]{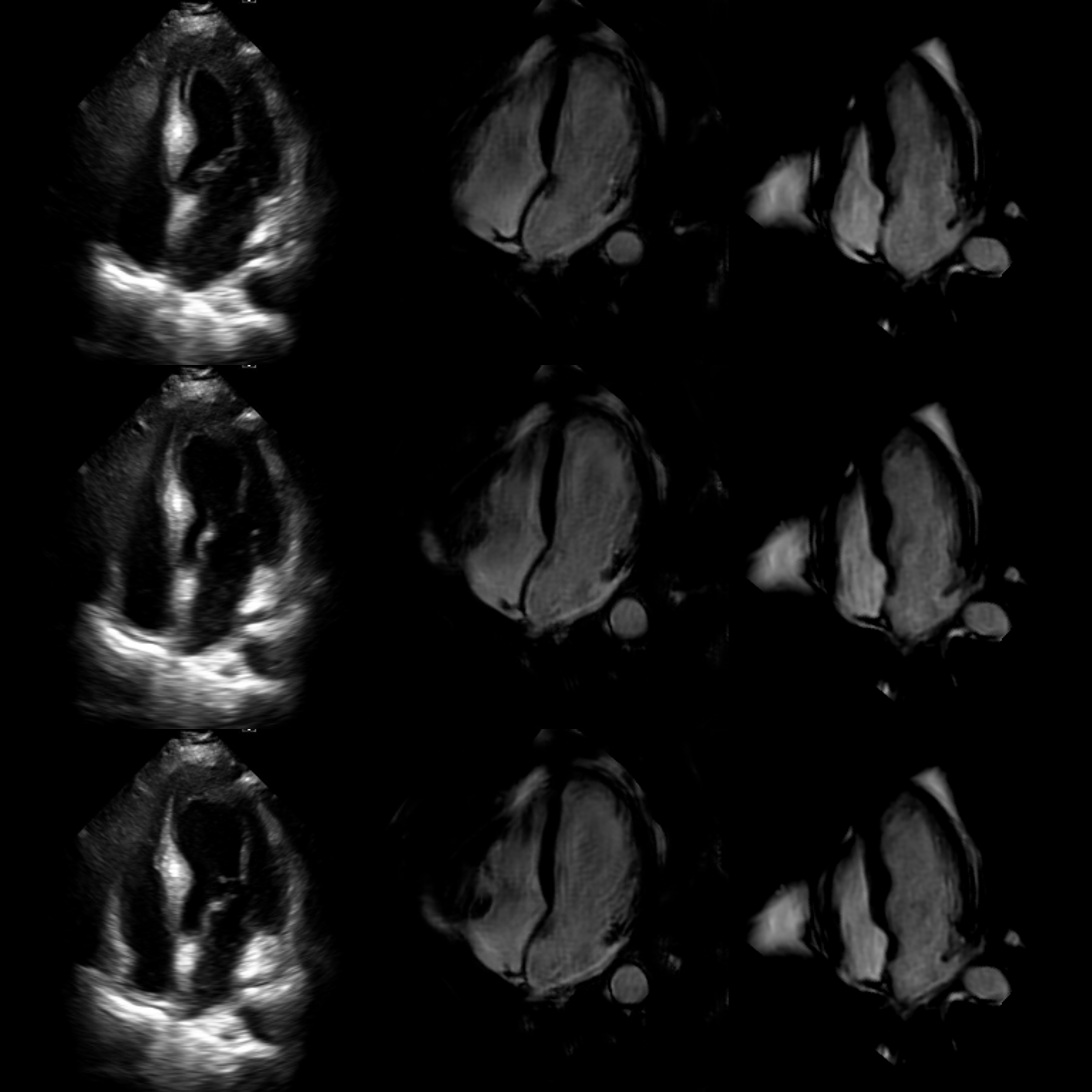}};
            \begin{scope}[x={(image.south east)}, y={(image.north west)}]
                \node[anchor=center, text=white] at (0.03, 0.97) {A};
                \node[anchor=center, text=white] at (0.37, 0.97) {B};
                \node[anchor=center, text=white] at (0.70, 0.97) {C};
            \end{scope}
        \end{tikzpicture}
        \captionof{figure}{\textbf{Patient 2:} Comparison of echocardiography (Column A), synthetic cardiac MRI (Column B), and real cardiac MRI (Column C) images. Echocardiography images show 3 different phases of one heartbeat, with corresponding synthetic and real cardiac MRI views.}
        \label{fig:AppendixPatient2}
    \end{minipage}
    \hfill
    \begin{minipage}[t]{0.49\textwidth}
        \centering
        \begin{tikzpicture}
            \node[anchor=south west, inner sep=0] (image) at (0,0) {%
              \includegraphics[width=\linewidth]{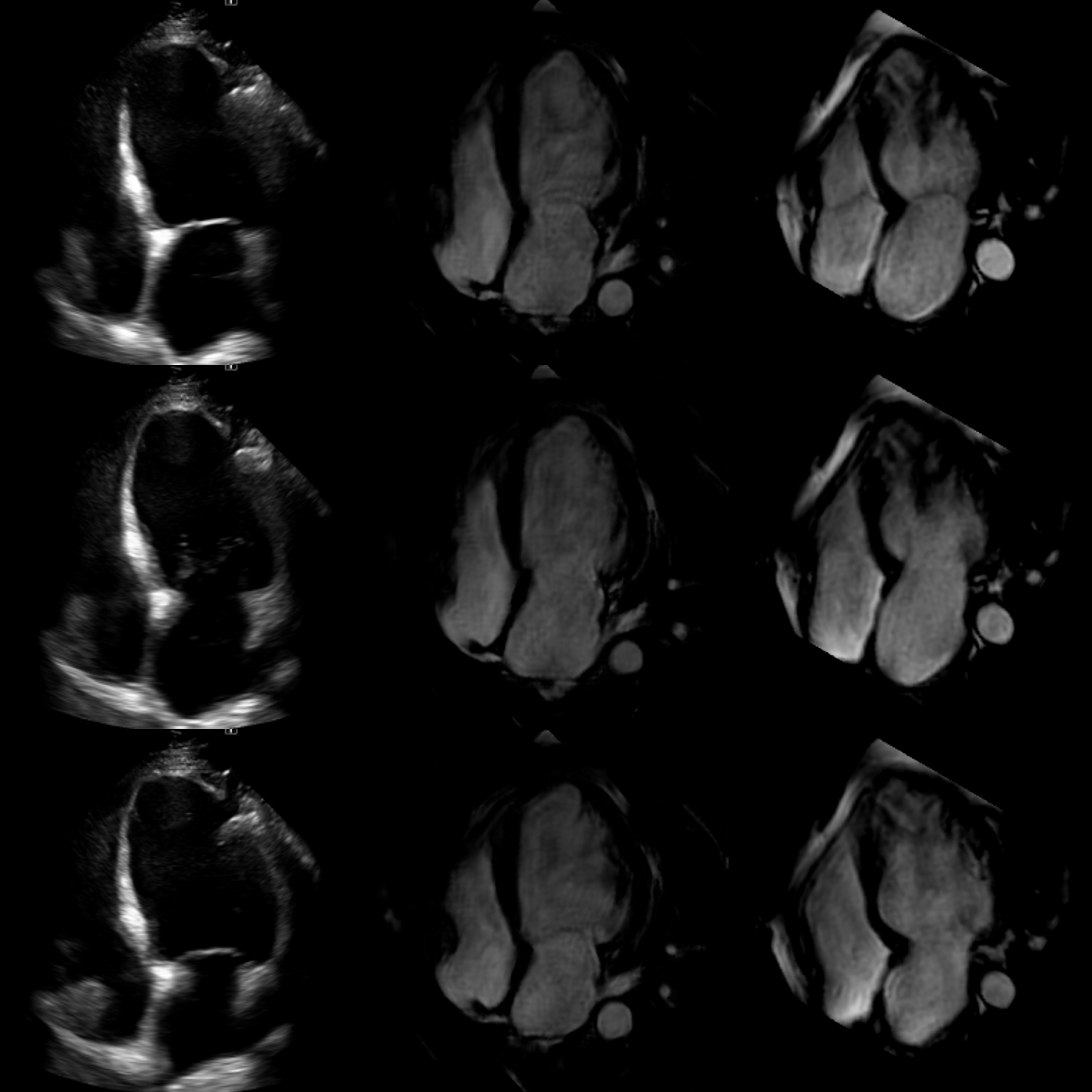}};
            \begin{scope}[x={(image.south east)}, y={(image.north west)}]
                \node[anchor=center, text=white] at (0.03, 0.97) {A};
                \node[anchor=center, text=white] at (0.37, 0.97) {B};
                \node[anchor=center, text=white] at (0.70, 0.97) {C};
            \end{scope}
        \end{tikzpicture}
        \captionof{figure}{\textbf{Patient 2:} Comparison of echocardiography (Column A), synthetic cardiac MRI (Column B), and real cardiac MRI (Column C) images. Echocardiography images show 3 different phases of one heartbeat, with corresponding synthetic and real cardiac MRI views.}
        \label{fig:AppendixPatient4}
    \end{minipage}
    
    \vspace{1em}
    
    \begin{minipage}[t]{0.49\textwidth}
        \centering
        \begin{tikzpicture}
            \node[anchor=south west, inner sep=0] (image) at (0,0) {%
              \includegraphics[width=\linewidth]{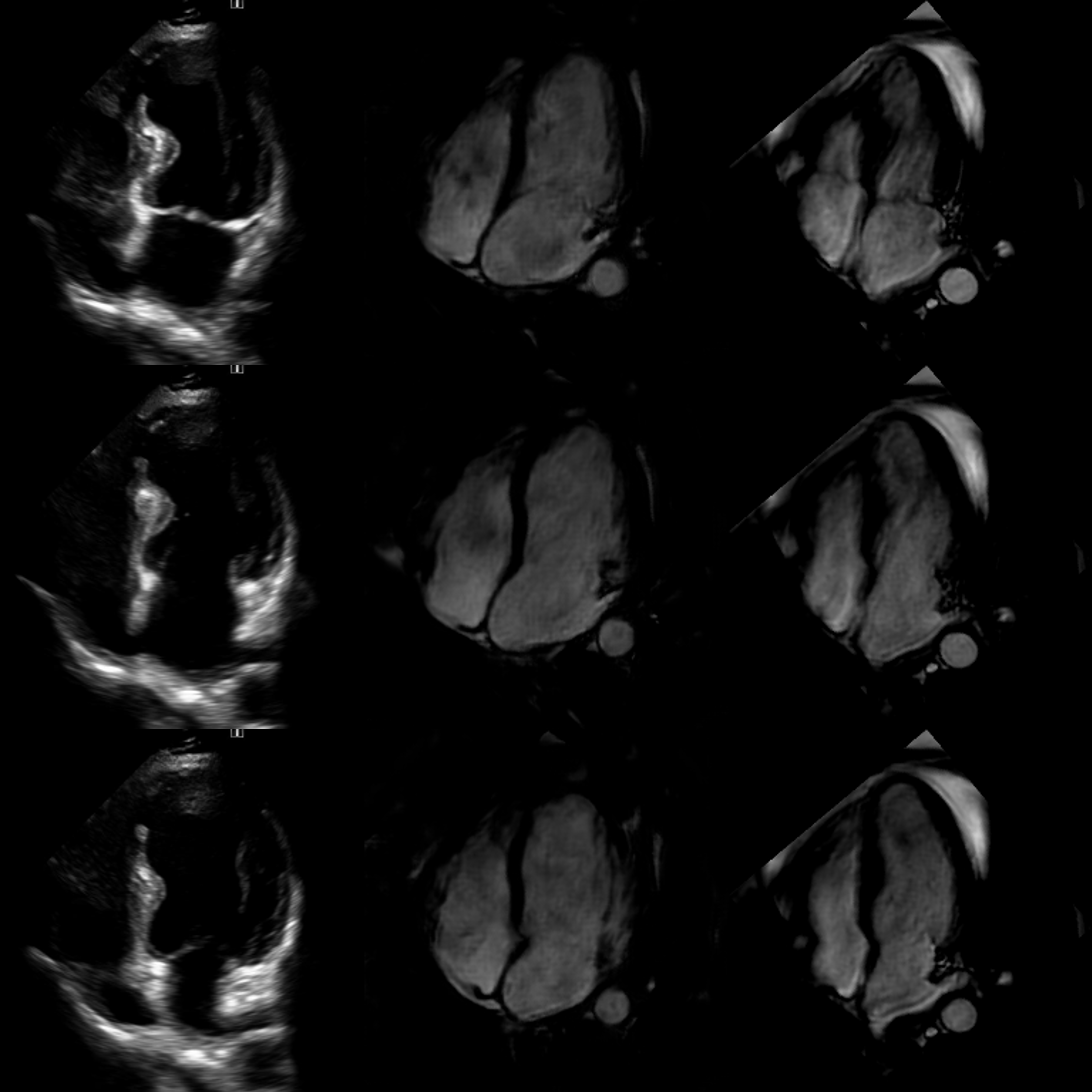}};
            \begin{scope}[x={(image.south east)}, y={(image.north west)}]
                \node[anchor=center, text=white] at (0.03, 0.97) {A};
                \node[anchor=center, text=white] at (0.37, 0.97) {B};
                \node[anchor=center, text=white] at (0.70, 0.97) {C};
            \end{scope}
        \end{tikzpicture}
        \captionof{figure}{\textbf{Patient 9:} Comparison of echocardiography (Column A), synthetic cardiac MRI (Column B), and real cardiac MRI (Column C) images. Echocardiography images show 3 different phases of one heartbeat, with corresponding synthetic and real cardiac MRI views.}
        \label{fig:AppendixPatient9}
    \end{minipage}
    \hfill
    \begin{minipage}[t]{0.49\textwidth}
        \centering
        \begin{tikzpicture}
            \node[anchor=south west, inner sep=0] (image) at (0,0) {%
              \includegraphics[width=\linewidth]{adali8_10.png}};
            \begin{scope}[x={(image.south east)}, y={(image.north west)}]
                \node[anchor=center, text=white] at (0.03, 0.97) {A};
                \node[anchor=center, text=white] at (0.37, 0.97) {B};
                \node[anchor=center, text=white] at (0.70, 0.97) {C};
            \end{scope}
        \end{tikzpicture}
        \captionof{figure}{\textbf{Patient 10:} Comparison of echocardiography (Column A), synthetic cardiac MRI (Column B), and real cardiac MRI (Column C) images. Echocardiography images show 3 different phases of one heartbeat, with corresponding synthetic and real cardiac MRI views.}
        \label{fig:AppendixPatient10}
    \end{minipage}
\end{figure}
\begin{figure}[!h]
    \centering
    \begin{minipage}[t]{0.49\textwidth}
        \centering
        \begin{tikzpicture}
            \node[anchor=south west, inner sep=0] (image) at (0,0) {%
              \includegraphics[width=\linewidth]{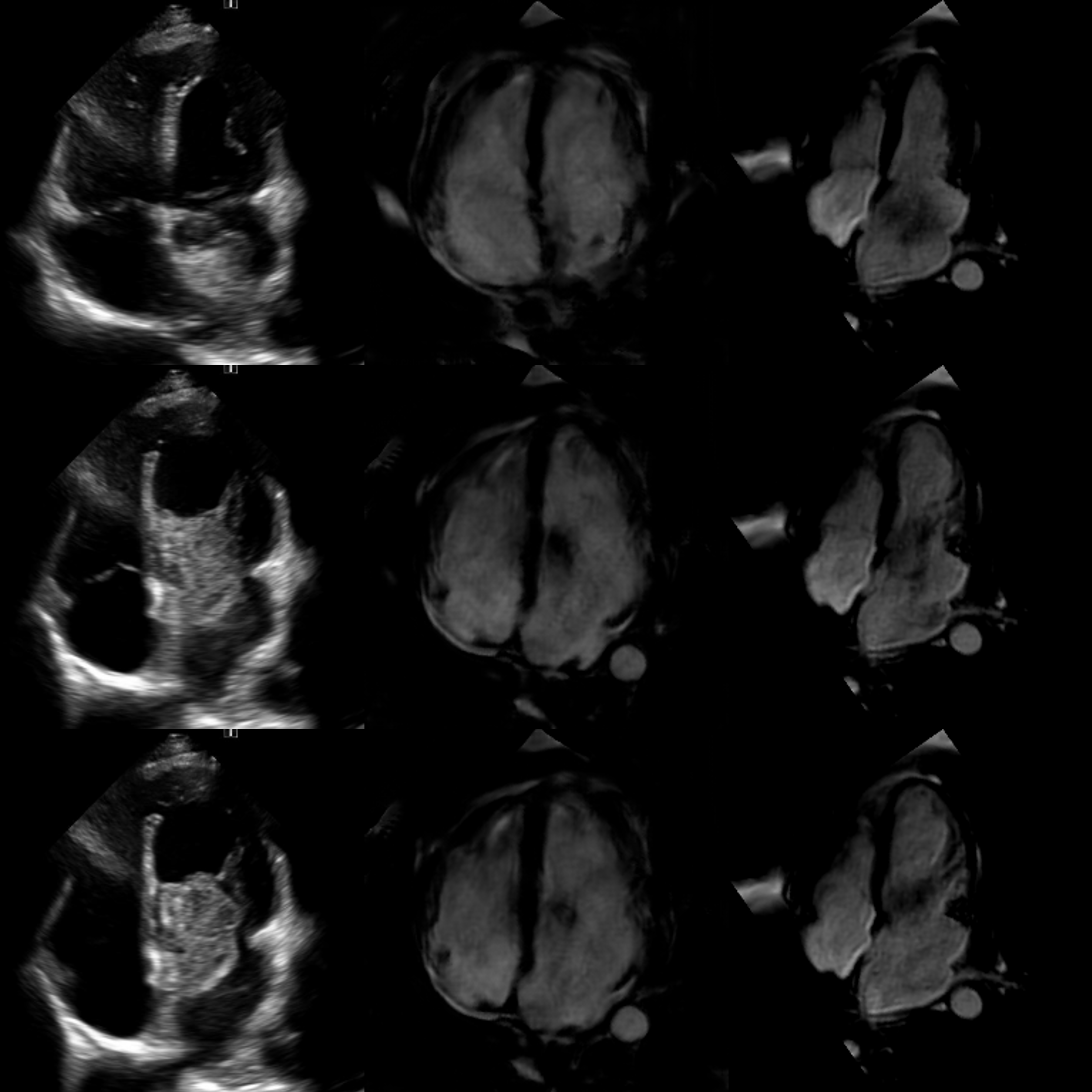}};
            \begin{scope}[x={(image.south east)}, y={(image.north west)}]
                \node[anchor=center, text=white] at (0.03, 0.97) {A};
                \node[anchor=center, text=white] at (0.37, 0.97) {B};
                \node[anchor=center, text=white] at (0.70, 0.97) {C};
            \end{scope}
        \end{tikzpicture}
        \captionof{figure}{\textbf{Patient 11:} Comparison of echocardiography (Column A), synthetic cardiac MRI (Column B), and real cardiac MRI (Column C) images. Echocardiography images show 3 different phases of one heartbeat, with corresponding synthetic and real cardiac MRI views.}
        \label{fig:AppendixPatient11}
    \end{minipage}
    \hfill
    \begin{minipage}[t]{0.49\textwidth}
        \centering
        \begin{tikzpicture}
            \node[anchor=south west, inner sep=0] (image) at (0,0) {%
              \includegraphics[width=\linewidth]{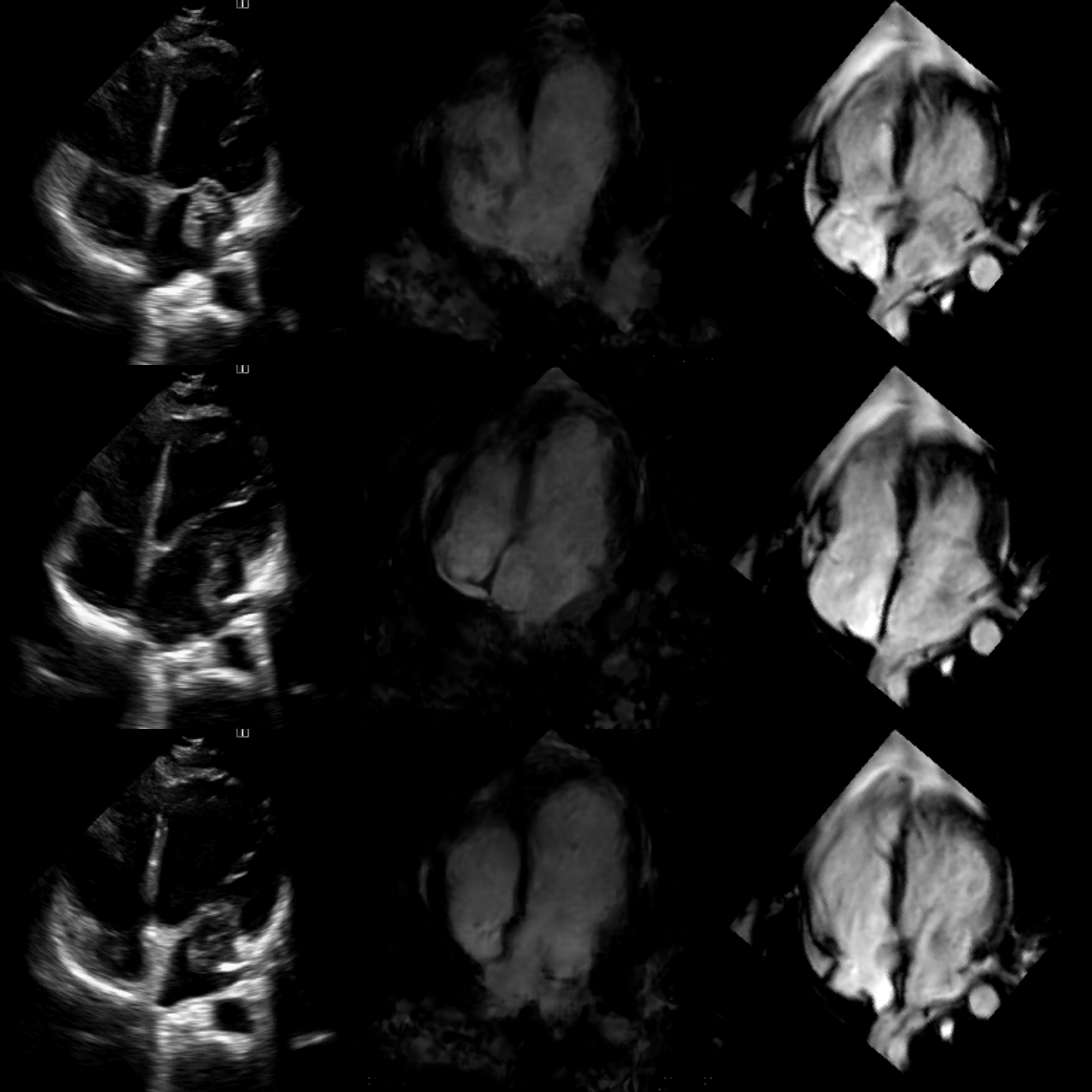}};
            \begin{scope}[x={(image.south east)}, y={(image.north west)}]
                \node[anchor=center, text=white] at (0.03, 0.97) {A};
                \node[anchor=center, text=white] at (0.37, 0.97) {B};
                \node[anchor=center, text=white] at (0.70, 0.97) {C};
            \end{scope}
        \end{tikzpicture}
        \captionof{figure}{\textbf{Patient 16:} Comparison of echocardiography (Column A), synthetic cardiac MRI (Column B), and real cardiac MRI (Column C) images. Echocardiography images show 3 different phases of one heartbeat, with corresponding synthetic and real cardiac MRI views.}
        \label{fig:AppendixPatient16}
    \end{minipage}
    
    \vspace{1em}
    
    \begin{minipage}[t]{0.49\textwidth}
        \centering
        \begin{tikzpicture}
            \node[anchor=south west, inner sep=0] (image) at (0,0) {%
              \includegraphics[width=\linewidth]{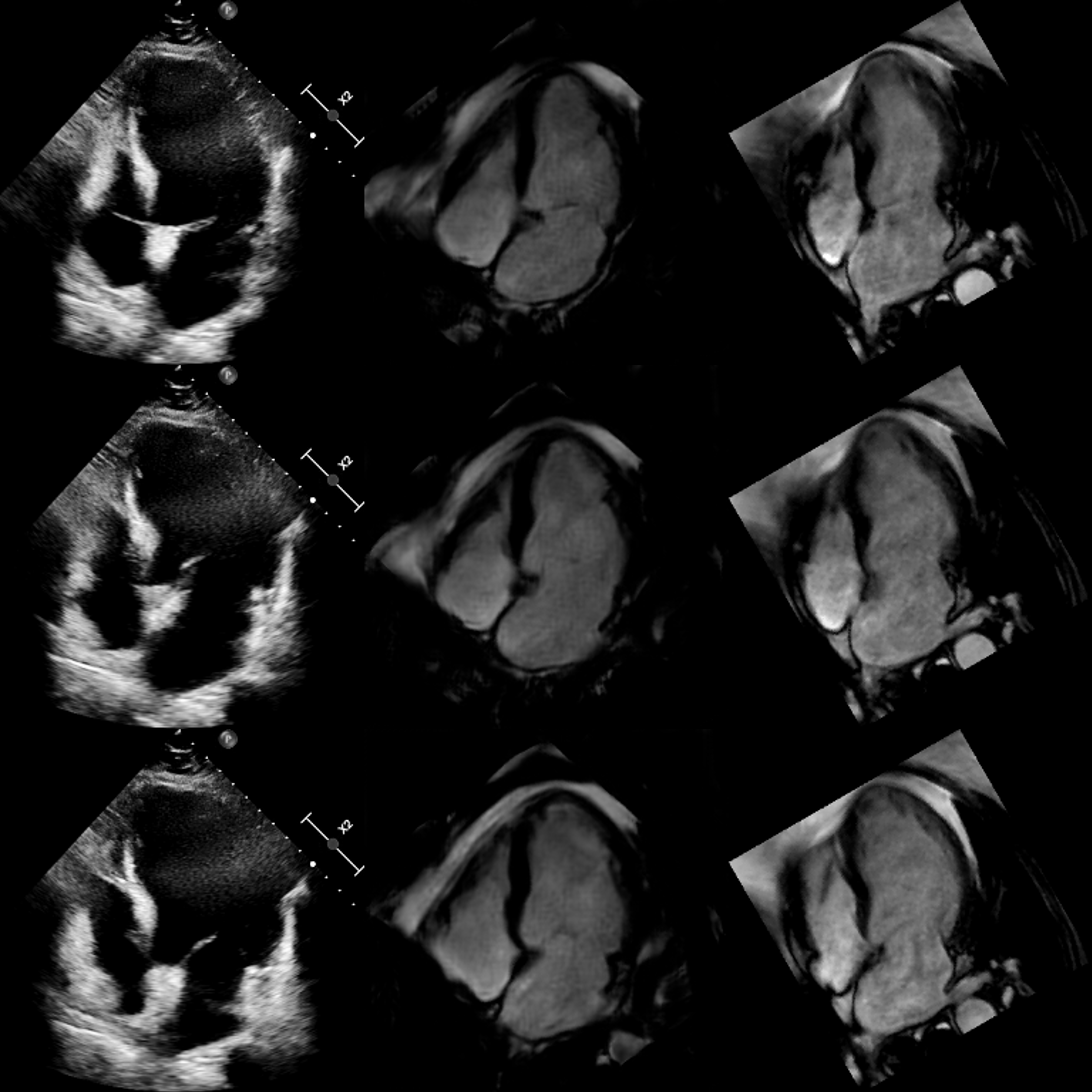}};
            \begin{scope}[x={(image.south east)}, y={(image.north west)}]
                \node[anchor=center, text=white] at (0.03, 0.97) {A};
                \node[anchor=center, text=white] at (0.37, 0.97) {B};
                \node[anchor=center, text=white] at (0.70, 0.97) {C};
            \end{scope}
        \end{tikzpicture}
        \captionof{figure}{\textbf{Patient 19:} Comparison of echocardiography (Column A), synthetic cardiac MRI (Column B), and real cardiac MRI (Column C) images. Echocardiography images show 3 different phases of one heartbeat, with corresponding synthetic and real cardiac MRI views.}
        \label{fig:AppendixPatient19}
    \end{minipage}
    \hfill
    \begin{minipage}[t]{0.49\textwidth}
        \centering
        \begin{tikzpicture}
            \node[anchor=south west, inner sep=0] (image) at (0,0) {%
              \includegraphics[width=\linewidth]{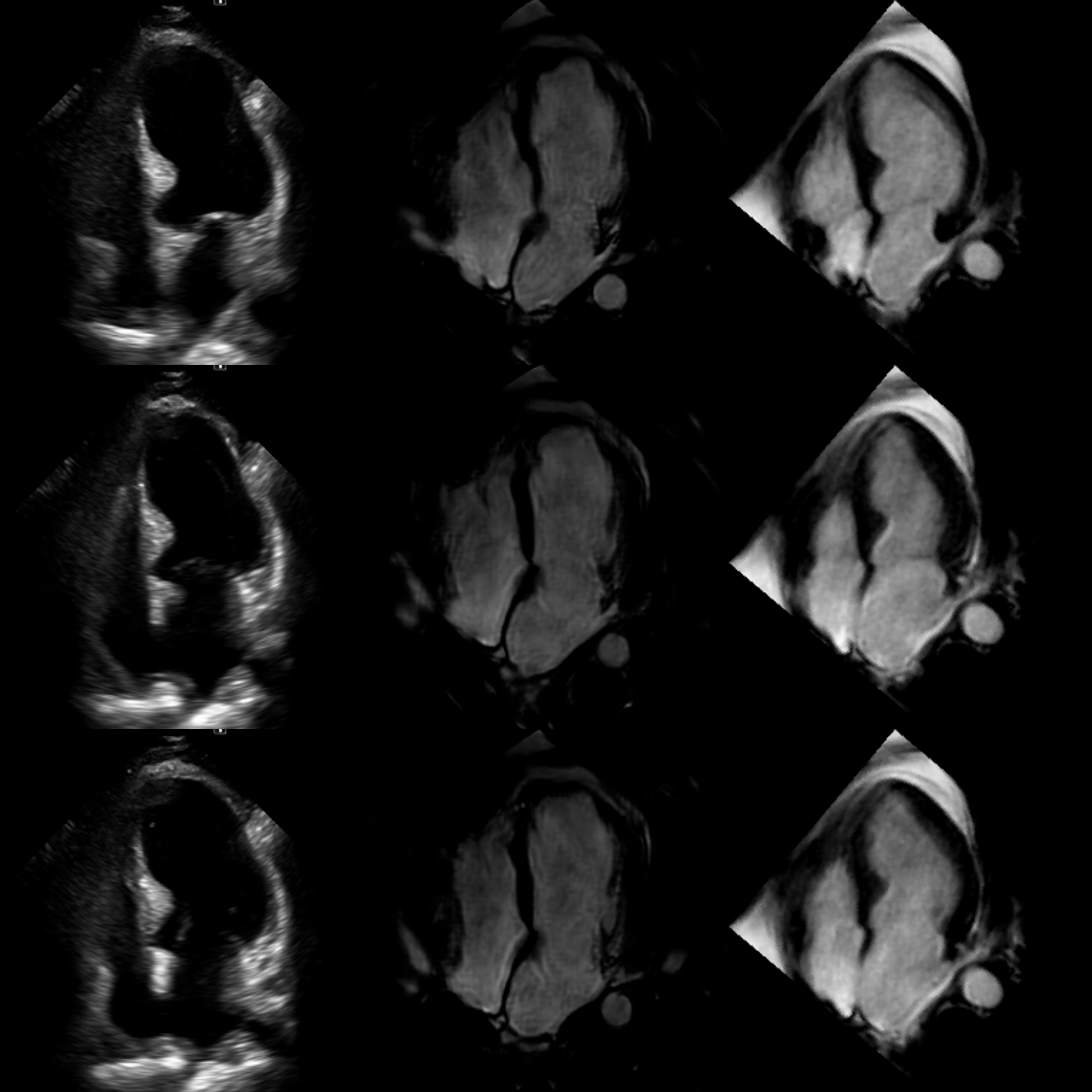}};
            \begin{scope}[x={(image.south east)}, y={(image.north west)}]
                \node[anchor=center, text=white] at (0.03, 0.97) {A};
                \node[anchor=center, text=white] at (0.37, 0.97) {B};
                \node[anchor=center, text=white] at (0.70, 0.97) {C};
            \end{scope}
        \end{tikzpicture}
        \captionof{figure}{\textbf{Patient 20:} Comparison of echocardiography (Column A), synthetic cardiac MRI (Column B), and real cardiac MRI (Column C) images. Echocardiography images show 3 different phases of one heartbeat, with corresponding synthetic and real cardiac MRI views.}
        \label{fig:AppendixPatient20}
    \end{minipage}
\end{figure}
\begin{figure}[!h]
    \centering
    \begin{minipage}[t]{0.49\textwidth}
        \centering
        \begin{tikzpicture}
            \node[anchor=south west, inner sep=0] (image) at (0,0) {%
              \includegraphics[width=\linewidth]{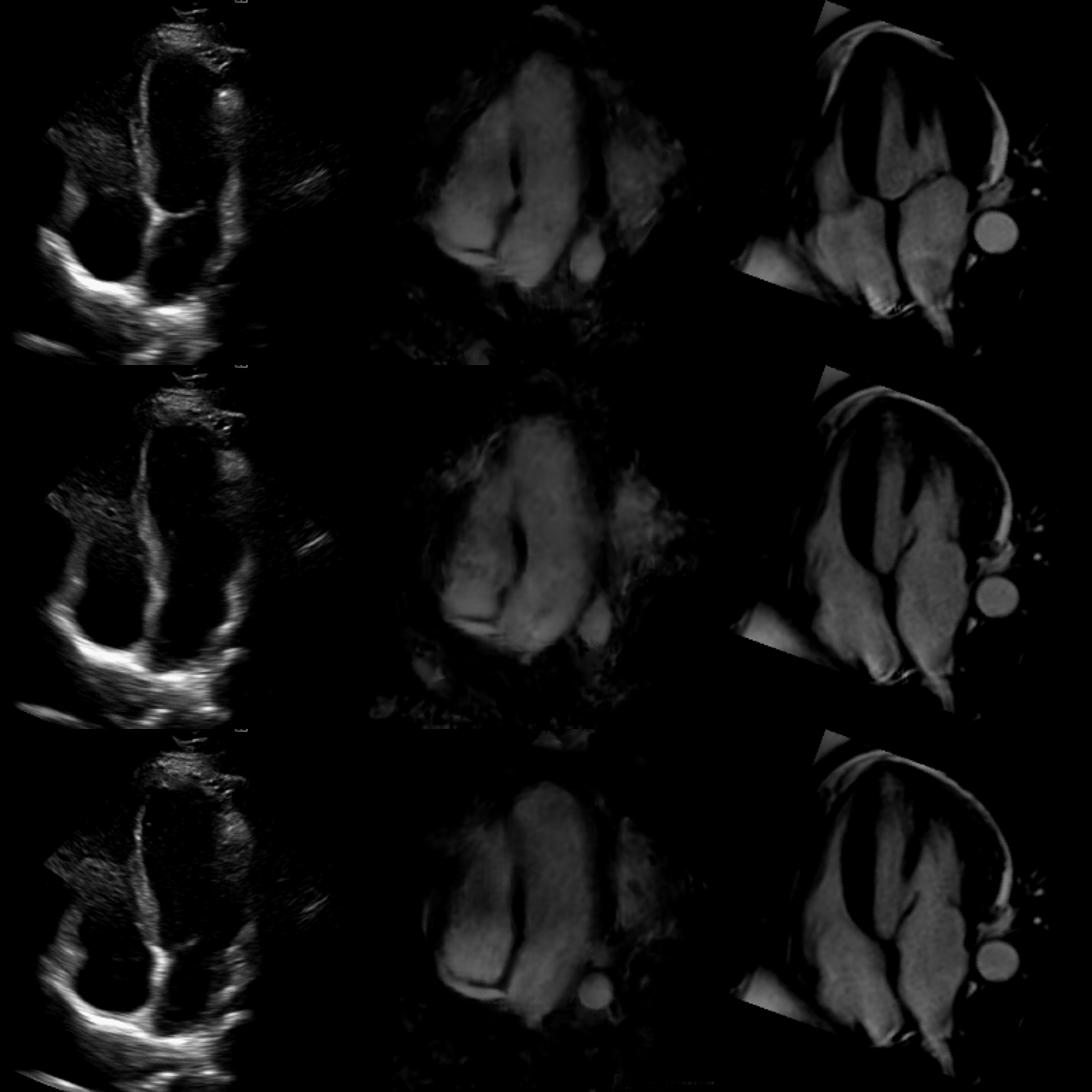}};
            \begin{scope}[x={(image.south east)}, y={(image.north west)}]
                \node[anchor=center, text=white] at (0.03, 0.97) {A};
                \node[anchor=center, text=white] at (0.37, 0.97) {B};
                \node[anchor=center, text=white] at (0.70, 0.97) {C};
            \end{scope}
        \end{tikzpicture}
        \captionof{figure}{\textbf{Patient 21:} Comparison of echocardiography (Column A), synthetic cardiac MRI (Column B), and real cardiac MRI (Column C) images. Echocardiography images show 3 different phases of one heartbeat, with corresponding synthetic and real cardiac MRI views.}
        \label{fig:AppendixPatient21}
    \end{minipage}
    \hfill
    \begin{minipage}[t]{0.49\textwidth}
        \centering
        \begin{tikzpicture}
            \node[anchor=south west, inner sep=0] (image) at (0,0) {%
              \includegraphics[width=\linewidth]{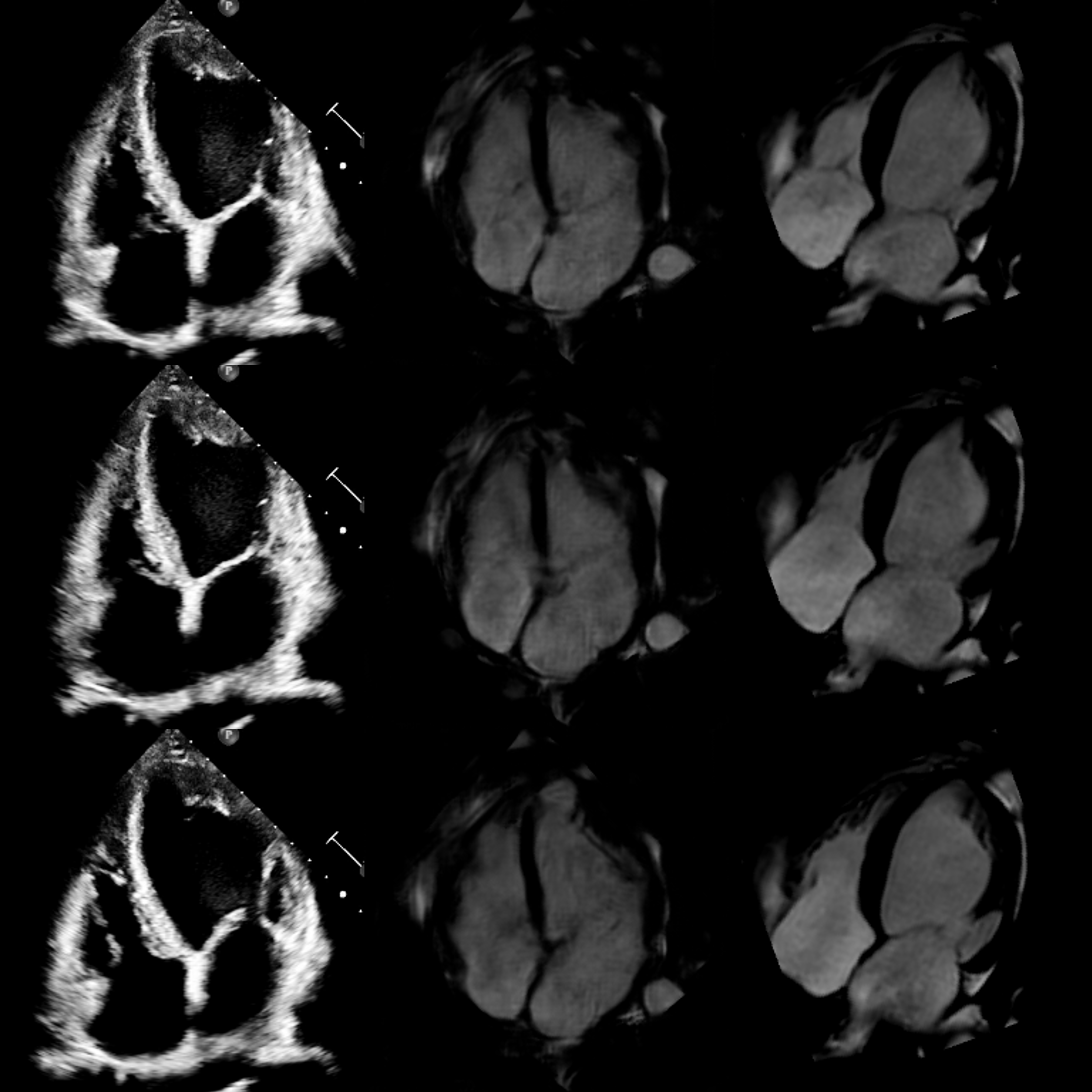}};
            \begin{scope}[x={(image.south east)}, y={(image.north west)}]
                \node[anchor=center, text=white] at (0.03, 0.97) {A};
                \node[anchor=center, text=white] at (0.37, 0.97) {B};
                \node[anchor=center, text=white] at (0.70, 0.97) {C};
            \end{scope}
        \end{tikzpicture}
        \captionof{figure}{\textbf{Patient 22:} Comparison of echocardiography (Column A), synthetic cardiac MRI (Column B), and real cardiac MRI (Column C) images. Echocardiography images show 3 different phases of one heartbeat, with corresponding synthetic and real cardiac MRI views.}
        \label{fig:AppendixPatient22}
    \end{minipage}

    \vspace{1em}

    \begin{minipage}[t]{0.49\textwidth}
        \centering
        \begin{tikzpicture}
            \node[anchor=south west, inner sep=0] (image) at (0,0) {%
              \includegraphics[width=\linewidth]{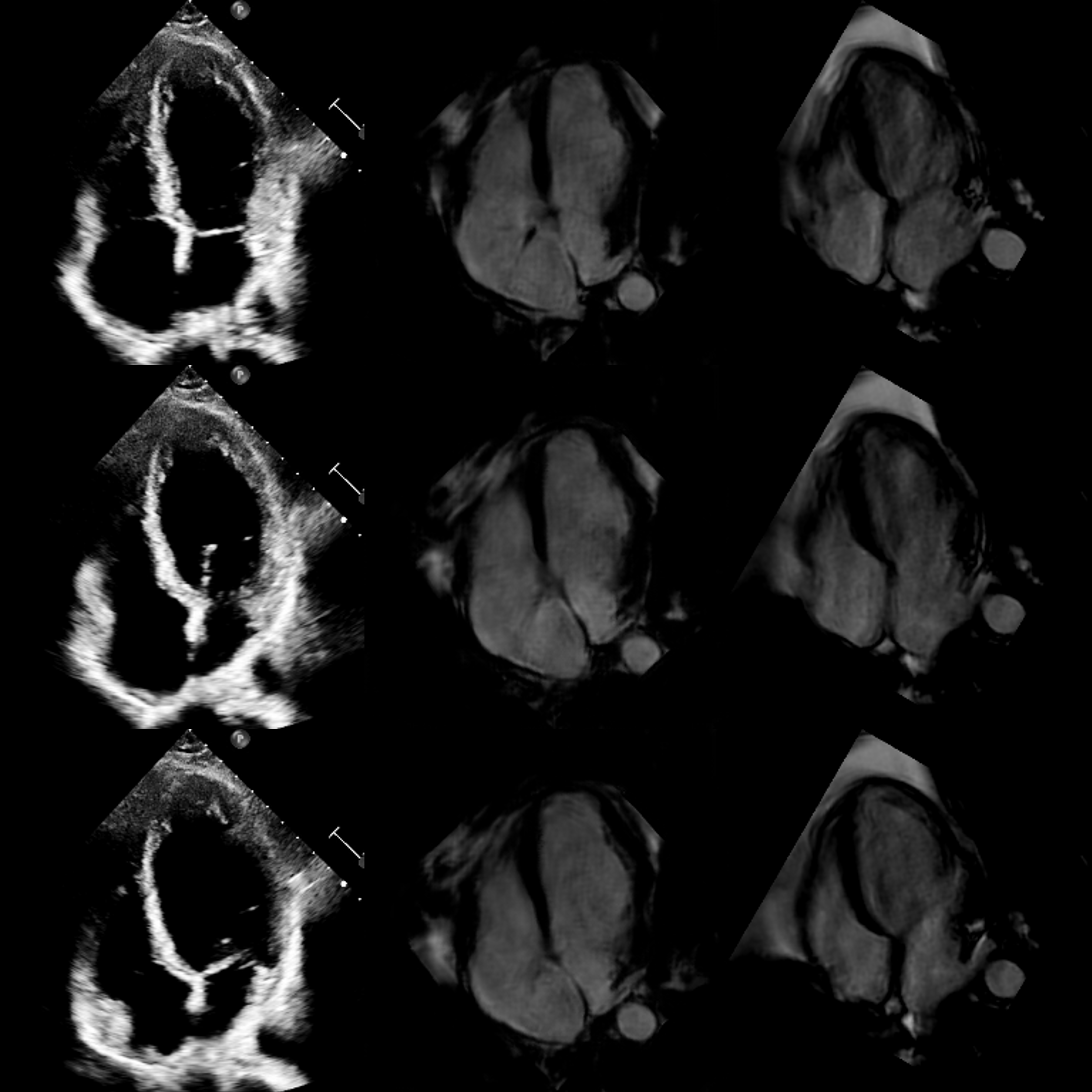}};
            \begin{scope}[x={(image.south east)}, y={(image.north west)}]
                \node[anchor=center, text=white] at (0.03, 0.97) {A};
                \node[anchor=center, text=white] at (0.37, 0.97) {B};
                \node[anchor=center, text=white] at (0.70, 0.97) {C};
            \end{scope}
        \end{tikzpicture}
        \captionof{figure}{\textbf{Patient 29:} Comparison of echocardiography (Column A), synthetic cardiac MRI (Column B), and real cardiac MRI (Column C) images. Echocardiography images show 3 different phases of one heartbeat, with corresponding synthetic and real cardiac MRI views.}
        \label{fig:AppendixPatient29}
    \end{minipage}
    \hfill
    \begin{minipage}[t]{0.49\textwidth}
        \centering
        \begin{tikzpicture}
            \node[anchor=south west, inner sep=0] (image) at (0,0) {%
              \includegraphics[width=\linewidth]{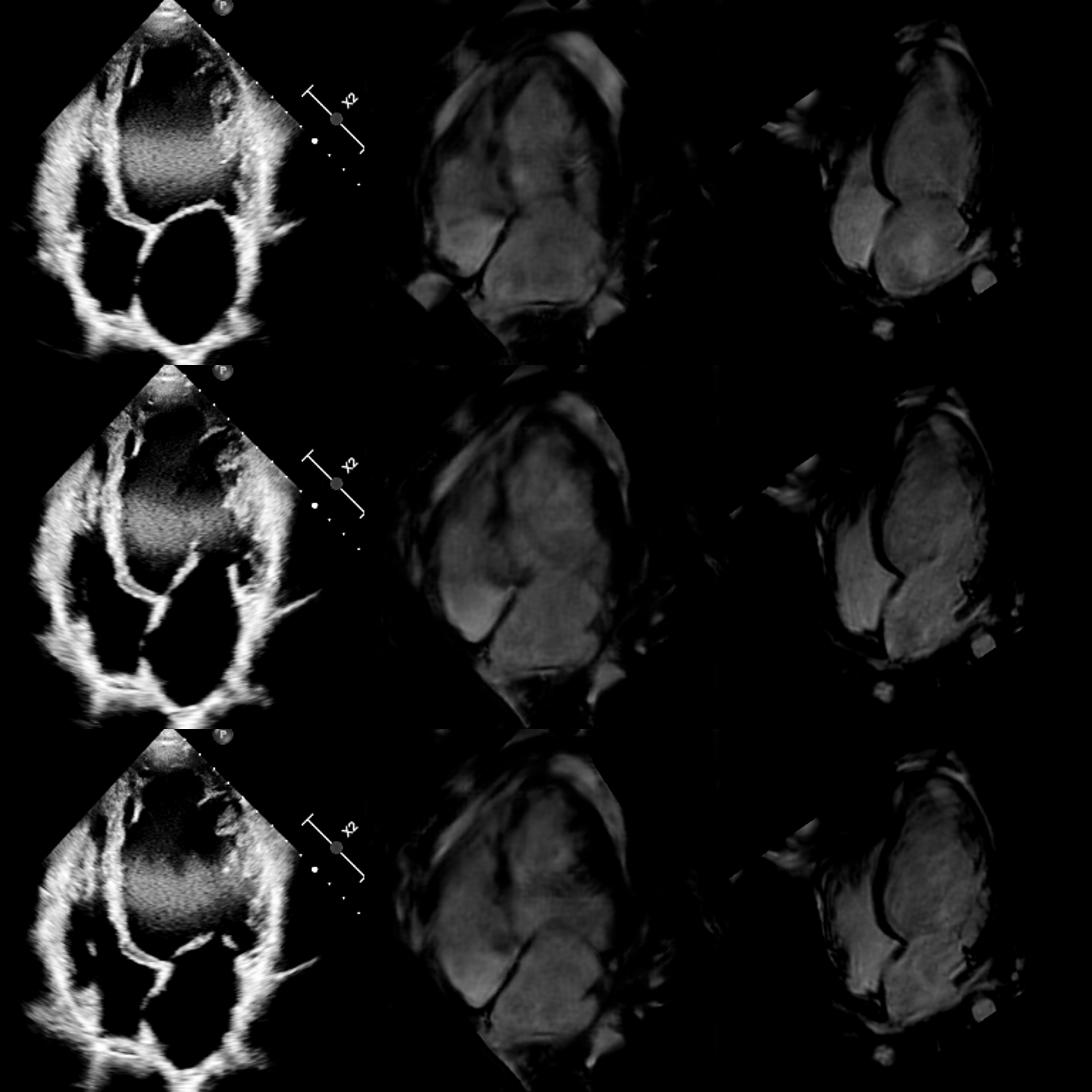}};
            \begin{scope}[x={(image.south east)}, y={(image.north west)}]
                \node[anchor=center, text=white] at (0.03, 0.97) {A};
                \node[anchor=center, text=white] at (0.37, 0.97) {B};
                \node[anchor=center, text=white] at (0.70, 0.97) {C};
            \end{scope}
        \end{tikzpicture}
        \captionof{figure}{\textbf{Patient 31:} Comparison of echocardiography (Column A), synthetic cardiac MRI (Column B), and real cardiac MRI (Column C) images. Echocardiography images show 3 different phases of one heartbeat, with corresponding synthetic and real cardiac MRI views.}
        \label{fig:AppendixPatient31}
    \end{minipage}
\end{figure}
\begin{figure}[!h]
    \centering
    \begin{minipage}[t]{0.49\textwidth}
        \centering
        \begin{tikzpicture}
            \node[anchor=south west, inner sep=0] (image) at (0,0) {%
              \includegraphics[width=\linewidth]{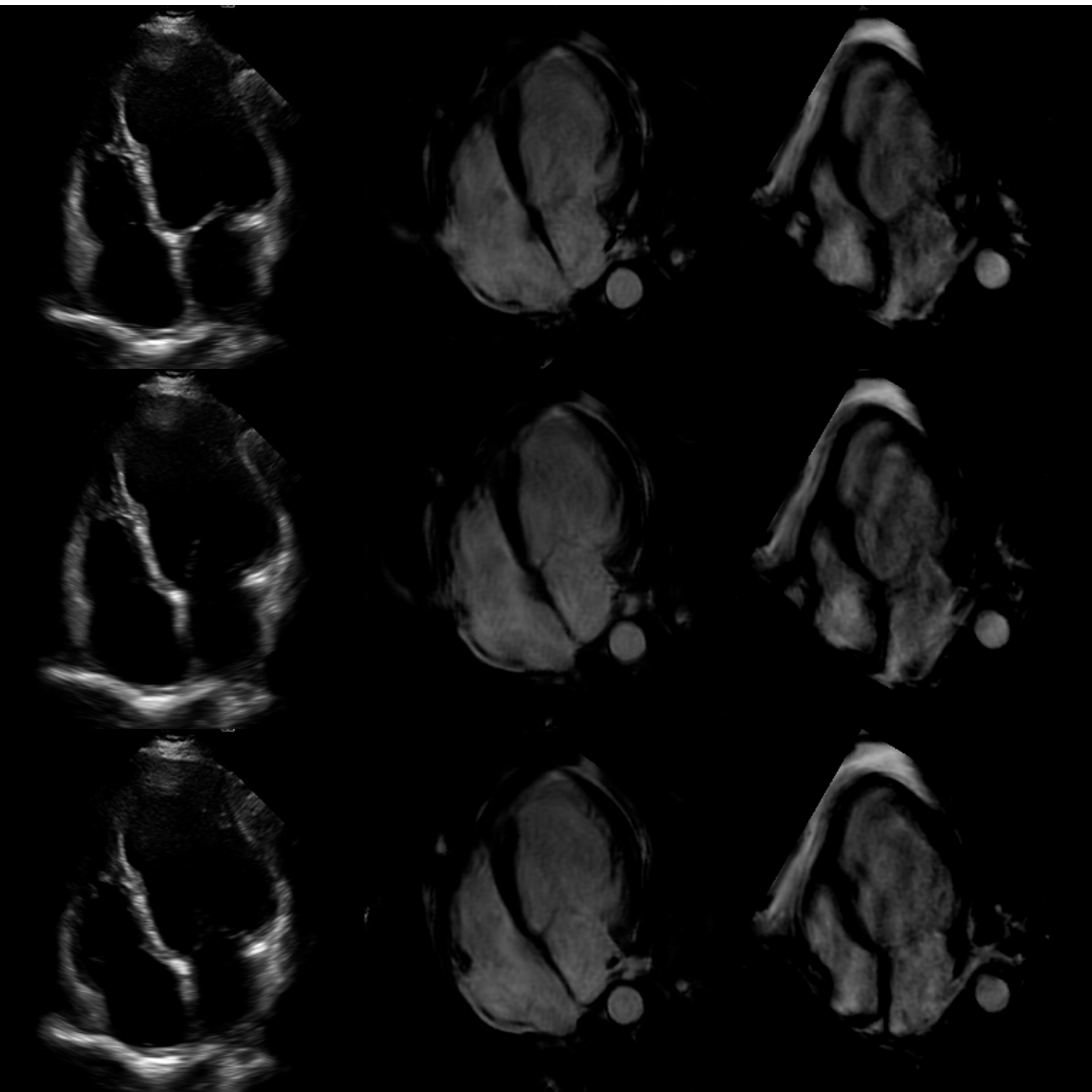}};
            \begin{scope}[x={(image.south east)}, y={(image.north west)}]
                \node[anchor=center, text=white] at (0.03, 0.97) {A};
                \node[anchor=center, text=white] at (0.37, 0.97) {B};
                \node[anchor=center, text=white] at (0.70, 0.97) {C};
            \end{scope}
        \end{tikzpicture}
        \captionof{figure}{\textbf{Patient 35:} Comparison of echocardiography (Column A), synthetic cardiac MRI (Column B), and real cardiac MRI (Column C) images. Echocardiography images show 3 different phases of one heartbeat, with corresponding synthetic and real cardiac MRI views.}
        \label{fig:AppendixPatient35}
    \end{minipage}
    \hfill
    \begin{minipage}[t]{0.49\textwidth}
        \centering
        \begin{tikzpicture}
            \node[anchor=south west, inner sep=0] (image) at (0,0) {%
              \includegraphics[width=\linewidth]{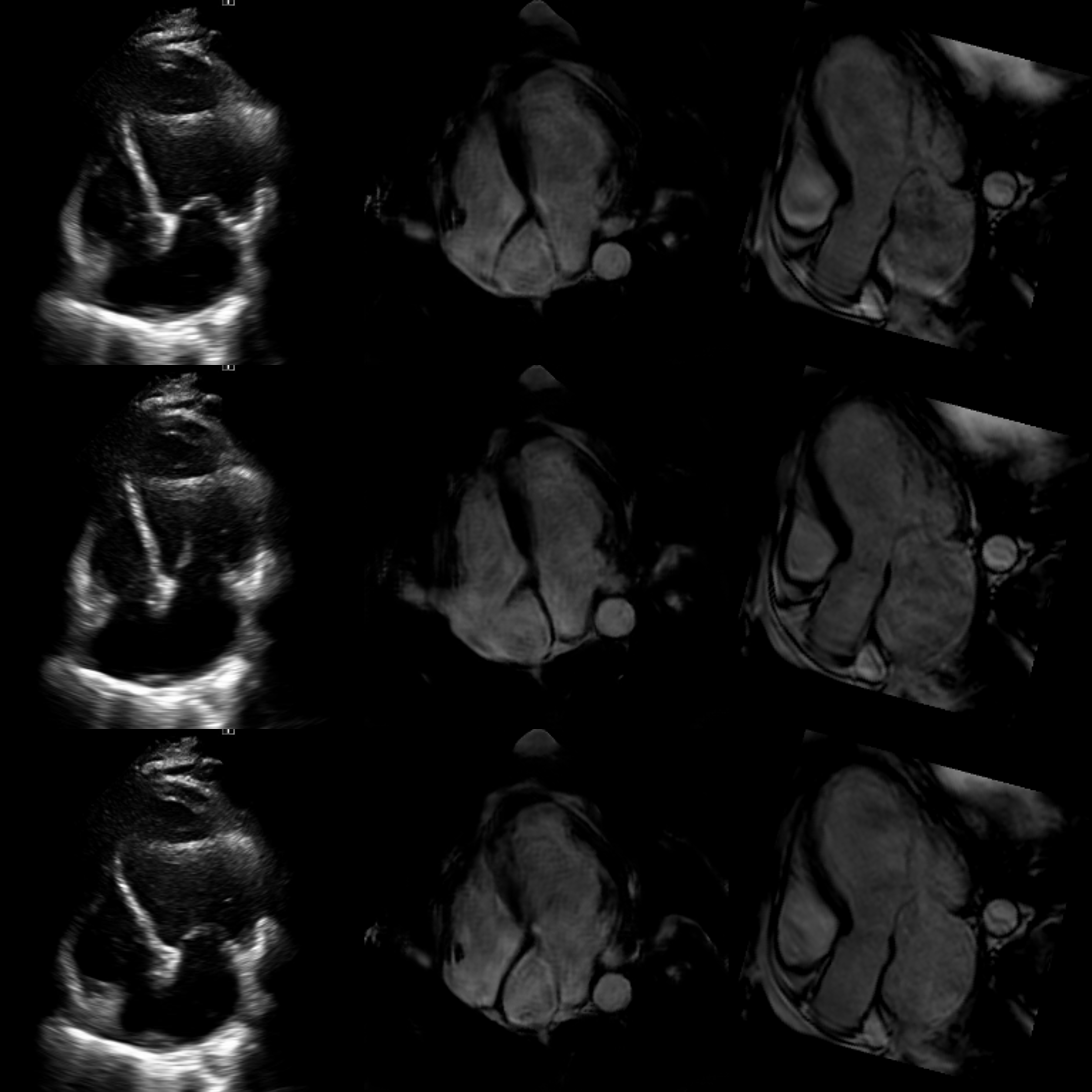}};
            \begin{scope}[x={(image.south east)}, y={(image.north west)}]
                \node[anchor=center, text=white] at (0.03, 0.97) {A};
                \node[anchor=center, text=white] at (0.37, 0.97) {B};
                \node[anchor=center, text=white] at (0.70, 0.97) {C};
            \end{scope}
        \end{tikzpicture}
        \captionof{figure}{\textbf{Patient 38:} Comparison of echocardiography (Column A), synthetic cardiac MRI (Column B), and real cardiac MRI (Column C) images. Echocardiography images show 3 different phases of one heartbeat, with corresponding synthetic and real cardiac MRI views.}
        \label{fig:AppendixPatient38}
    \end{minipage}
    
    \vspace{1em}
    
    \begin{minipage}[t]{0.49\textwidth}
        \centering
        \begin{tikzpicture}
            \node[anchor=south west, inner sep=0] (image) at (0,0) {%
              \includegraphics[width=\linewidth]{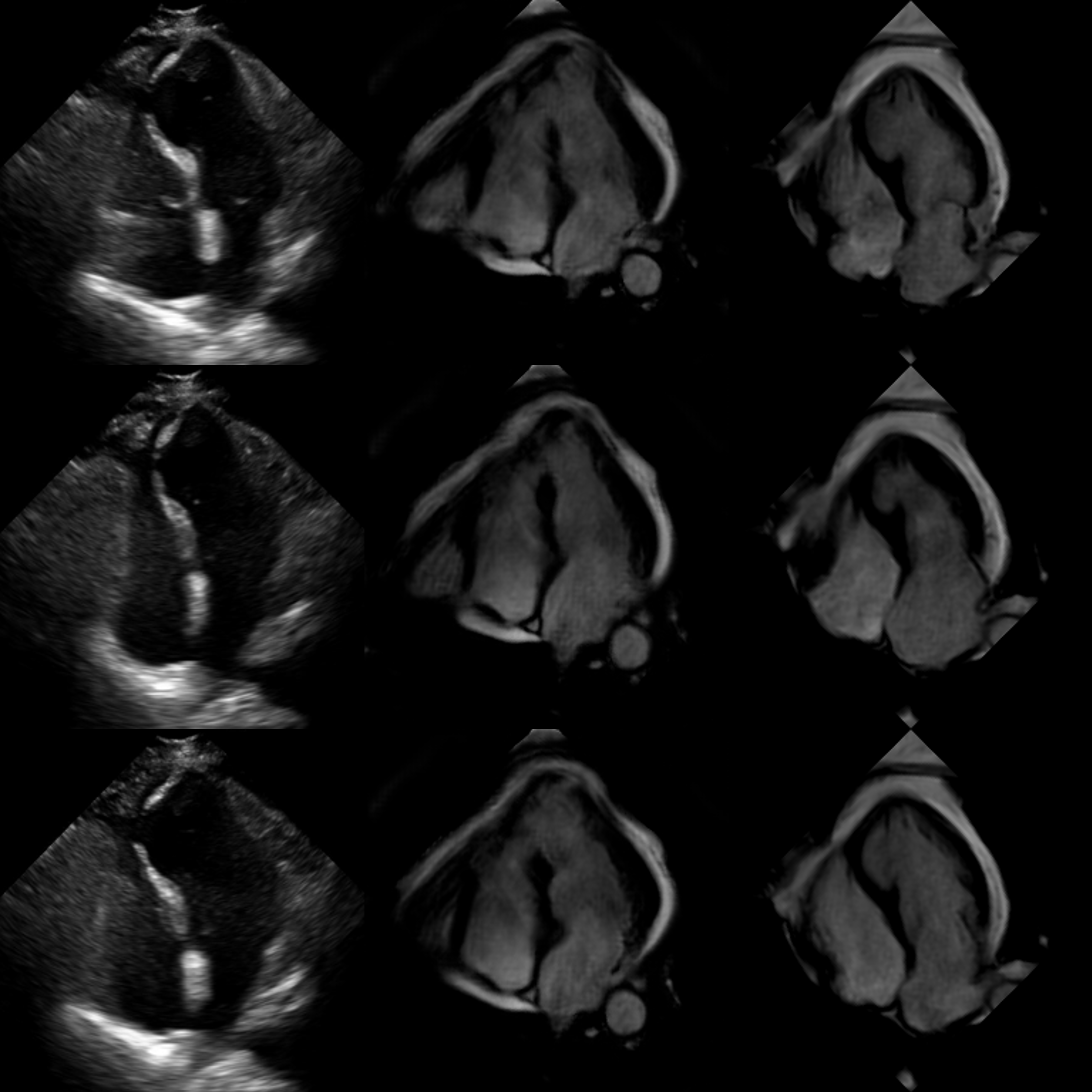}};
            \begin{scope}[x={(image.south east)}, y={(image.north west)}]
                \node[anchor=center, text=white] at (0.03, 0.97) {A};
                \node[anchor=center, text=white] at (0.37, 0.97) {B};
                \node[anchor=center, text=white] at (0.70, 0.97) {C};
            \end{scope}
        \end{tikzpicture}
        \captionof{figure}{\textbf{Patient 40:} Comparison of echocardiography (Column A), synthetic cardiac MRI (Column B), and real cardiac MRI (Column C) images. Echocardiography images show 3 different phases of one heartbeat, with corresponding synthetic and real cardiac MRI views.}
        \label{fig:AppendixPatient40}
    \end{minipage}
    \hfill
    \begin{minipage}[t]{0.49\textwidth}
        \centering
        \begin{tikzpicture}
            \node[anchor=south west, inner sep=0] (image) at (0,0) {%
              \includegraphics[width=\linewidth]{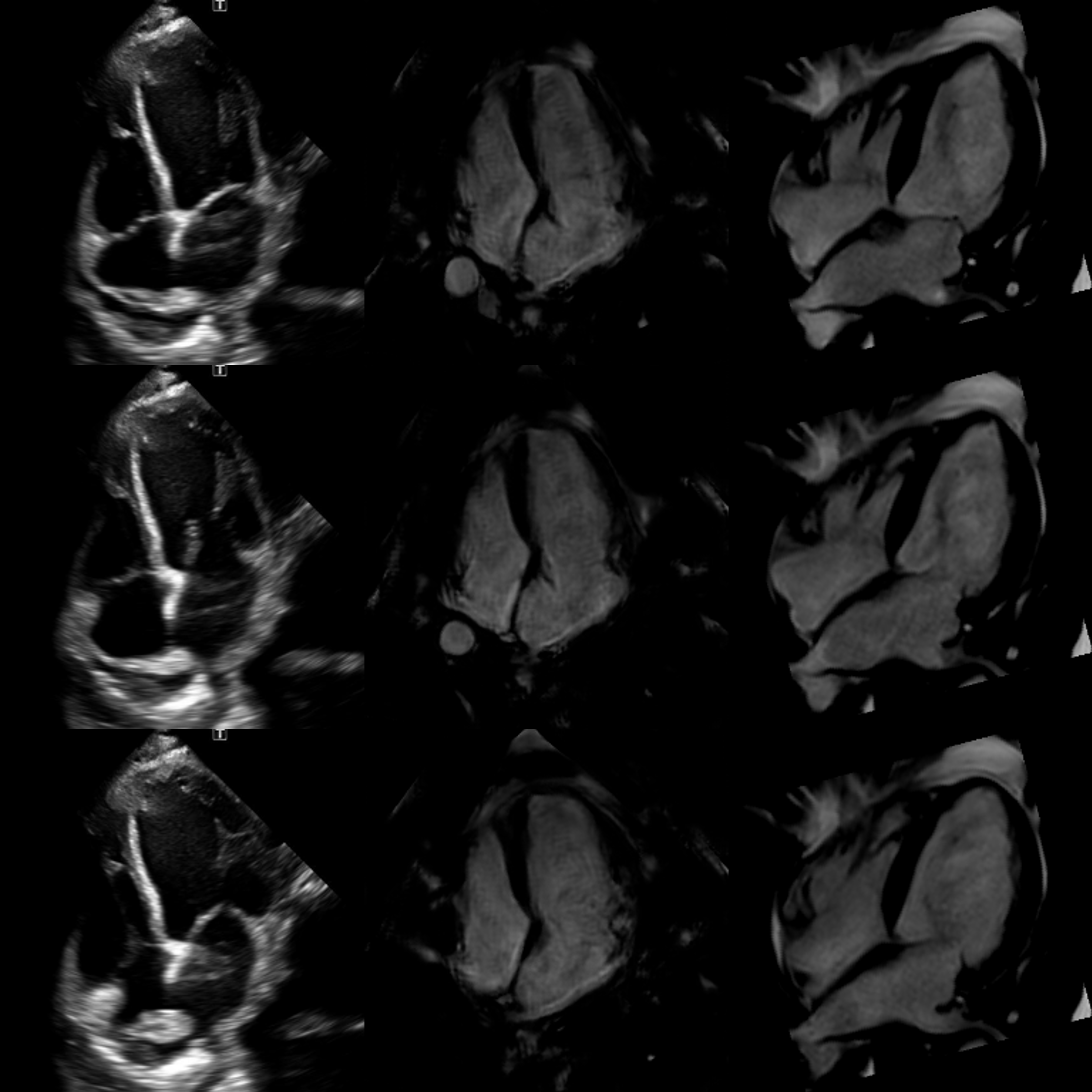}};
            \begin{scope}[x={(image.south east)}, y={(image.north west)}]
                \node[anchor=center, text=white] at (0.03, 0.97) {A};
                \node[anchor=center, text=white] at (0.37, 0.97) {B};
                \node[anchor=center, text=white] at (0.70, 0.97) {C};
            \end{scope}
        \end{tikzpicture}
        \captionof{figure}{\textbf{Patient 41:} Comparison of echocardiography (Column A), synthetic cardiac MRI (Column B), and real cardiac MRI (Column C) images. Echocardiography images show 3 different phases of one heartbeat, with corresponding synthetic and real cardiac MRI views.}
        \label{fig:AppendixPatient41}
    \end{minipage}
\end{figure}
\begin{figure}[!h]
    \centering
    \begin{minipage}[t]{0.49\textwidth}
        \centering
        \begin{tikzpicture}
            \node[anchor=south west, inner sep=0] (image) at (0,0) {%
              \includegraphics[width=\linewidth]{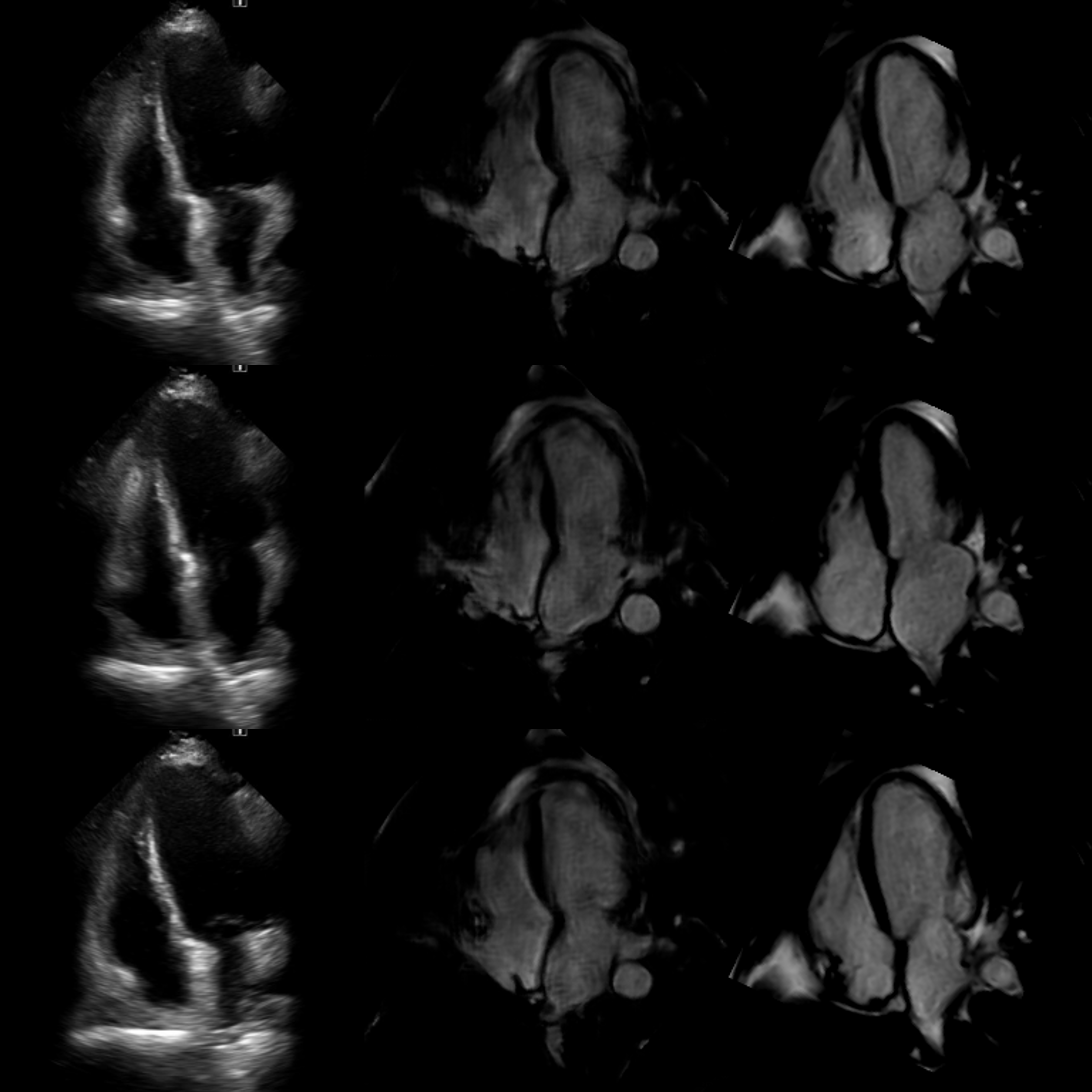}};
            \begin{scope}[x={(image.south east)}, y={(image.north west)}]
                \node[anchor=center, text=white] at (0.03, 0.97) {A};
                \node[anchor=center, text=white] at (0.37, 0.97) {B};
                \node[anchor=center, text=white] at (0.70, 0.97) {C};
            \end{scope}
        \end{tikzpicture}
        \captionof{figure}{\textbf{Patient 44:} Comparison of echocardiography (Column A), synthetic cardiac MRI (Column B), and real cardiac MRI (Column C) images. Echocardiography images show 3 different phases of one heartbeat, with corresponding synthetic and real cardiac MRI views.}
        \label{fig:AppendixPatient44}
    \end{minipage}
    \hfill
    \begin{minipage}[t]{0.49\textwidth}
        \centering
        \begin{tikzpicture}
            \node[anchor=south west, inner sep=0] (image) at (0,0) {%
              \includegraphics[width=\linewidth]{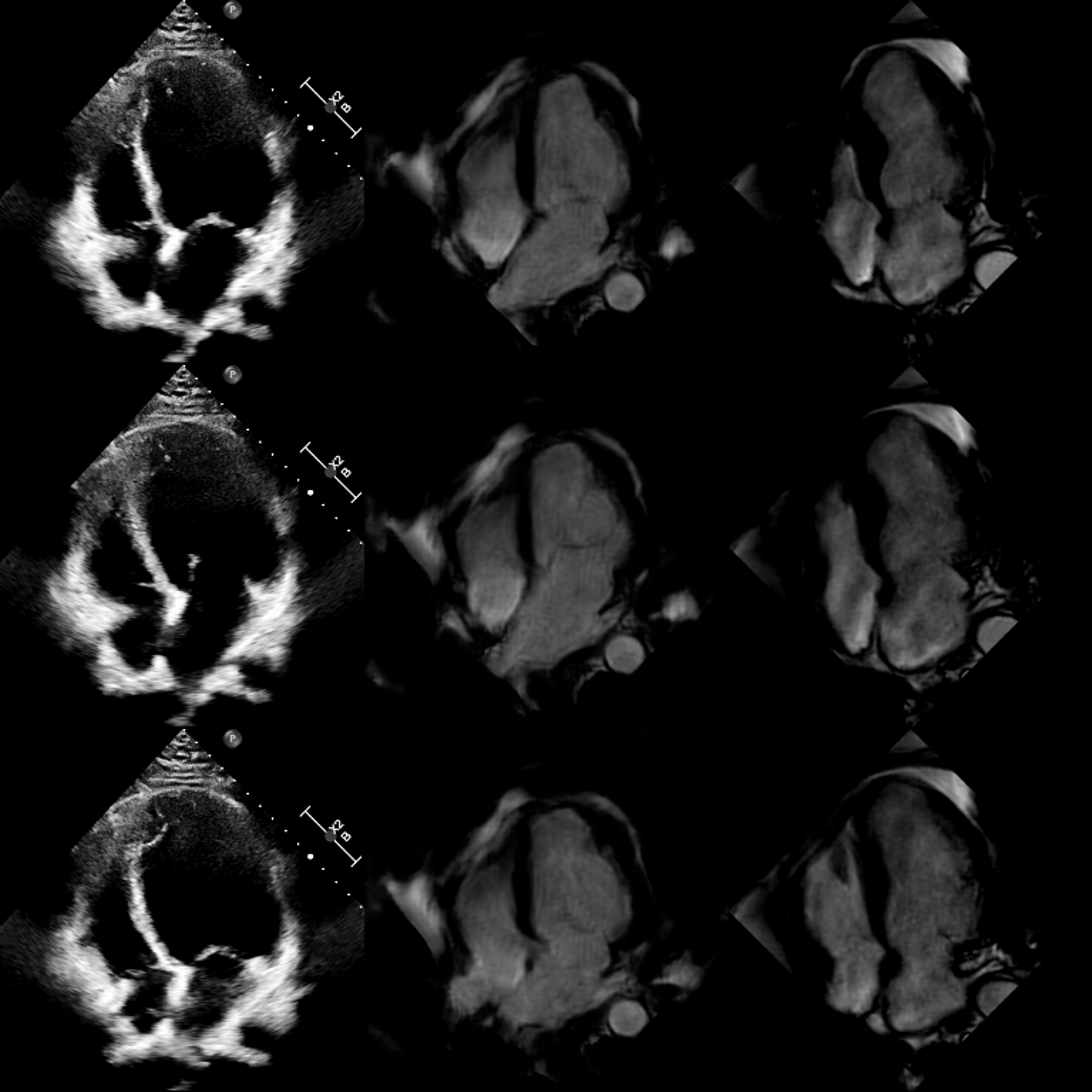}};
            \begin{scope}[x={(image.south east)}, y={(image.north west)}]
                \node[anchor=center, text=white] at (0.03, 0.97) {A};
                \node[anchor=center, text=white] at (0.37, 0.97) {B};
                \node[anchor=center, text=white] at (0.70, 0.97) {C};
            \end{scope}
        \end{tikzpicture}
        \captionof{figure}{\textbf{Patient 46:} Comparison of echocardiography (Column A), synthetic cardiac MRI (Column B), and real cardiac MRI (Column C) images. Echocardiography images show 3 different phases of one heartbeat, with corresponding synthetic and real cardiac MRI views.}
        \label{fig:AppendixPatient46}
    \end{minipage}
    
    \vspace{1em}
    
    \begin{minipage}[t]{0.49\textwidth}
        \centering
        \begin{tikzpicture}
            \node[anchor=south west, inner sep=0] (image) at (0,0) {%
              \includegraphics[width=\linewidth]{adali2_51.png}};
            \begin{scope}[x={(image.south east)}, y={(image.north west)}]
                \node[anchor=center, text=white] at (0.03, 0.97) {A};
                \node[anchor=center, text=white] at (0.37, 0.97) {B};
                \node[anchor=center, text=white] at (0.70, 0.97) {C};
            \end{scope}
        \end{tikzpicture}
        \captionof{figure}{\textbf{Patient 51:} Comparison of echocardiography (Column A), synthetic cardiac MRI (Column B), and real cardiac MRI (Column C) images. Echocardiography images show 3 different phases of one heartbeat, with corresponding synthetic and real cardiac MRI views.}
        \label{fig:AppendixPatient51}
    \end{minipage}
    \hfill
    \begin{minipage}[t]{0.49\textwidth}
        \centering
        \begin{tikzpicture}
            \node[anchor=south west, inner sep=0] (image) at (0,0) {%
              \includegraphics[width=\linewidth]{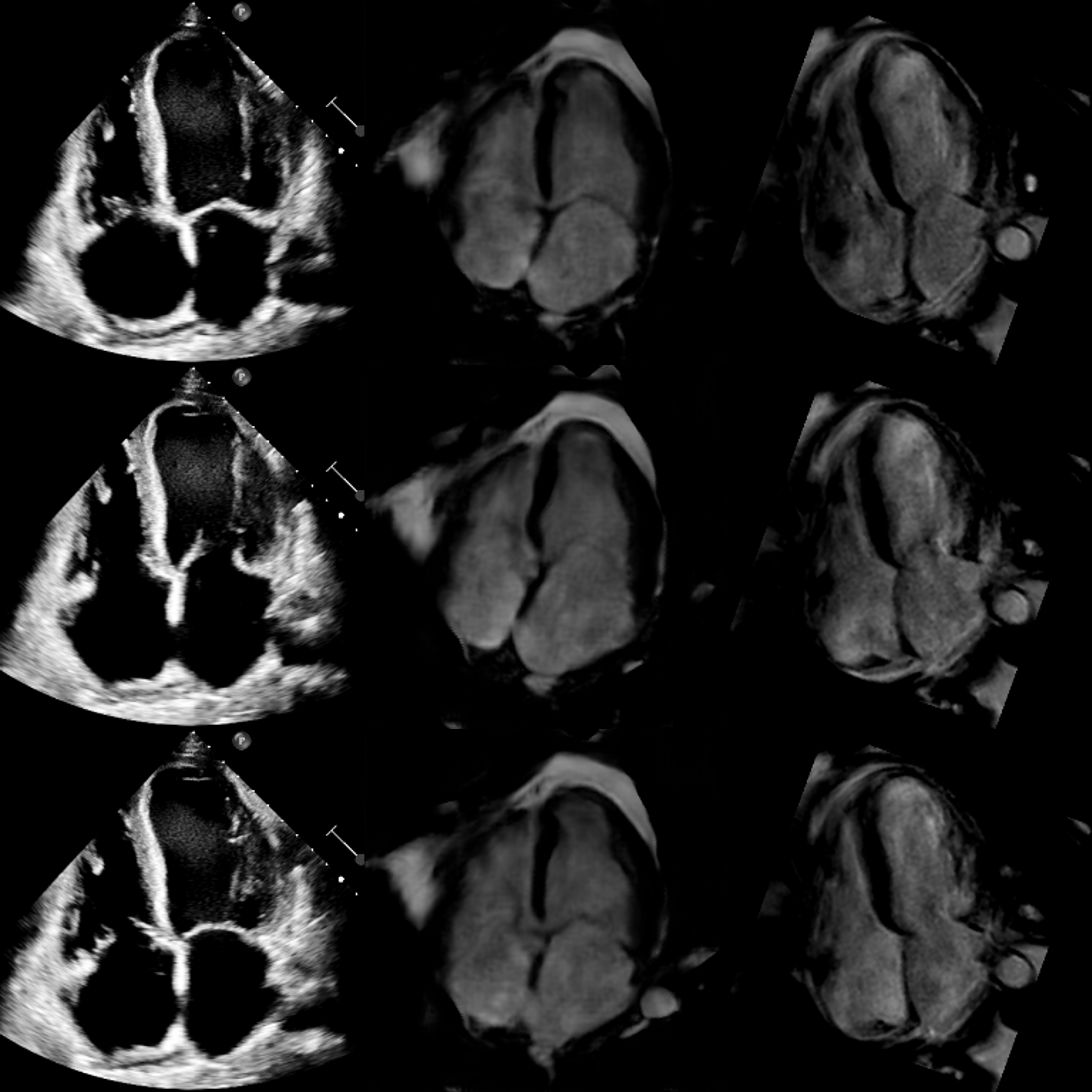}};
            \begin{scope}[x={(image.south east)}, y={(image.north west)}]
                \node[anchor=center, text=white] at (0.03, 0.97) {A};
                \node[anchor=center, text=white] at (0.37, 0.97) {B};
                \node[anchor=center, text=white] at (0.70, 0.97) {C};
            \end{scope}
        \end{tikzpicture}
        \captionof{figure}{\textbf{Patient 55:} Comparison of echocardiography (Column A), synthetic cardiac MRI (Column B), and real cardiac MRI (Column C) images. Echocardiography images show 3 different phases of one heartbeat, with corresponding synthetic and real cardiac MRI views.}
        \label{fig:AppendixPatient55}
    \end{minipage}
\end{figure}
\begin{figure}[!h]
    \centering
    \begin{minipage}[t]{0.49\textwidth}
        \centering
        \begin{tikzpicture}
            \node[anchor=south west, inner sep=0] (image) at (0,0) {%
              \includegraphics[width=\linewidth]{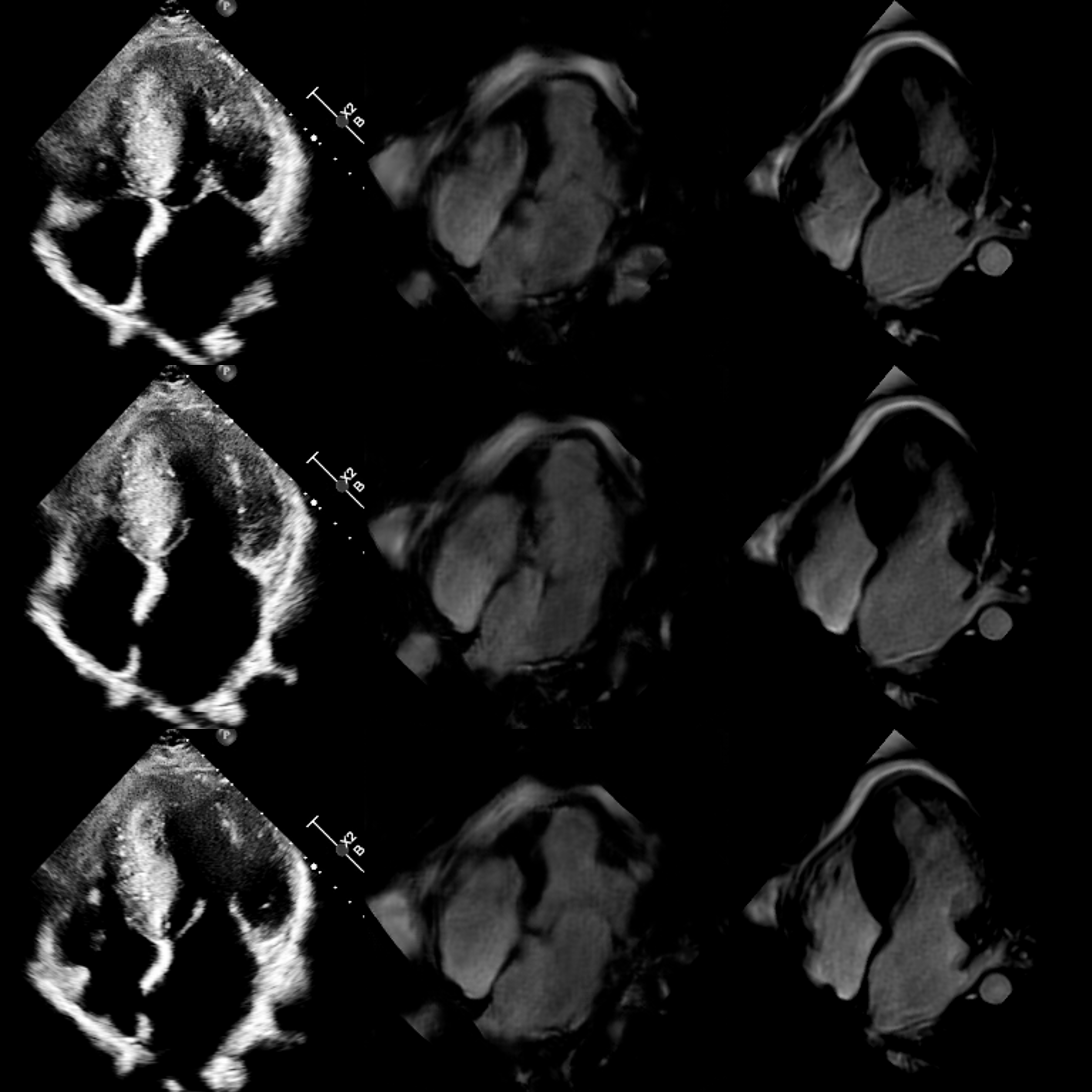}};
            \begin{scope}[x={(image.south east)}, y={(image.north west)}]
                \node[anchor=center, text=white] at (0.03, 0.97) {A};
                \node[anchor=center, text=white] at (0.37, 0.97) {B};
                \node[anchor=center, text=white] at (0.70, 0.97) {C};
            \end{scope}
        \end{tikzpicture}
        \captionof{figure}{\textbf{Patient 56:} Comparison of echocardiography (Column A), synthetic cardiac MRI (Column B), and real cardiac MRI (Column C) images. Echocardiography images show 3 different phases of one heartbeat, with corresponding synthetic and real cardiac MRI views.}
        \label{fig:AppendixPatient56}
    \end{minipage}
    \hfill
    \begin{minipage}[t]{0.49\textwidth}
        \centering
        \begin{tikzpicture}
            \node[anchor=south west, inner sep=0] (image) at (0,0) {%
              \includegraphics[width=\linewidth]{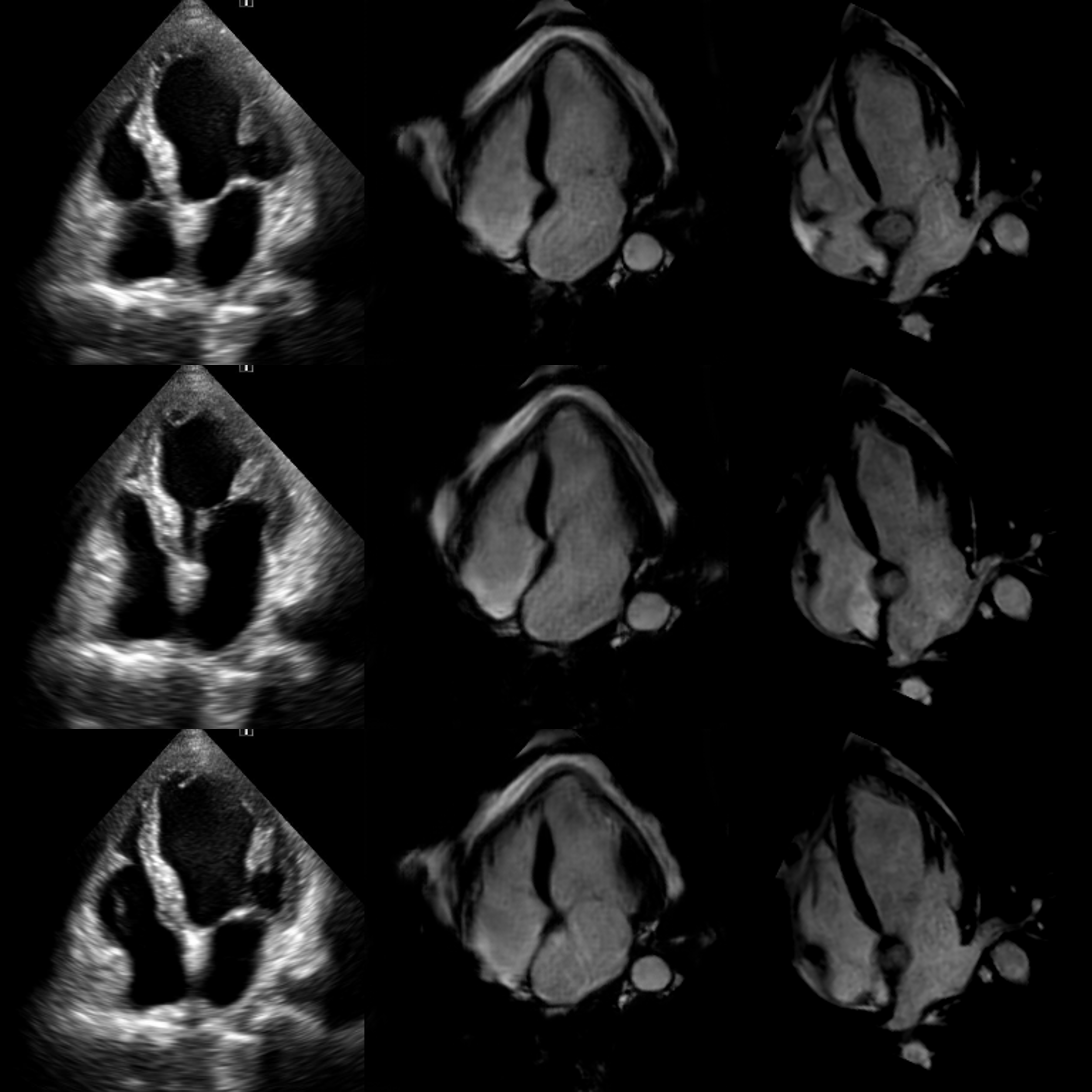}};
            \begin{scope}[x={(image.south east)}, y={(image.north west)}]
                \node[anchor=center, text=white] at (0.03, 0.97) {A};
                \node[anchor=center, text=white] at (0.37, 0.97) {B};
                \node[anchor=center, text=white] at (0.70, 0.97) {C};
            \end{scope}
        \end{tikzpicture}
        \captionof{figure}{\textbf{Patient 63:} Comparison of echocardiography (Column A), synthetic cardiac MRI (Column B), and real cardiac MRI (Column C) images. Echocardiography images show 3 different phases of one heartbeat, with corresponding synthetic and real cardiac MRI views.}
        \label{fig:AppendixPatient63}
    \end{minipage}
    
    \vspace{1em}
    
    \begin{minipage}[t]{0.49\textwidth}
        \centering
        \begin{tikzpicture}
            \node[anchor=south west, inner sep=0] (image) at (0,0) {%
              \includegraphics[width=\linewidth]{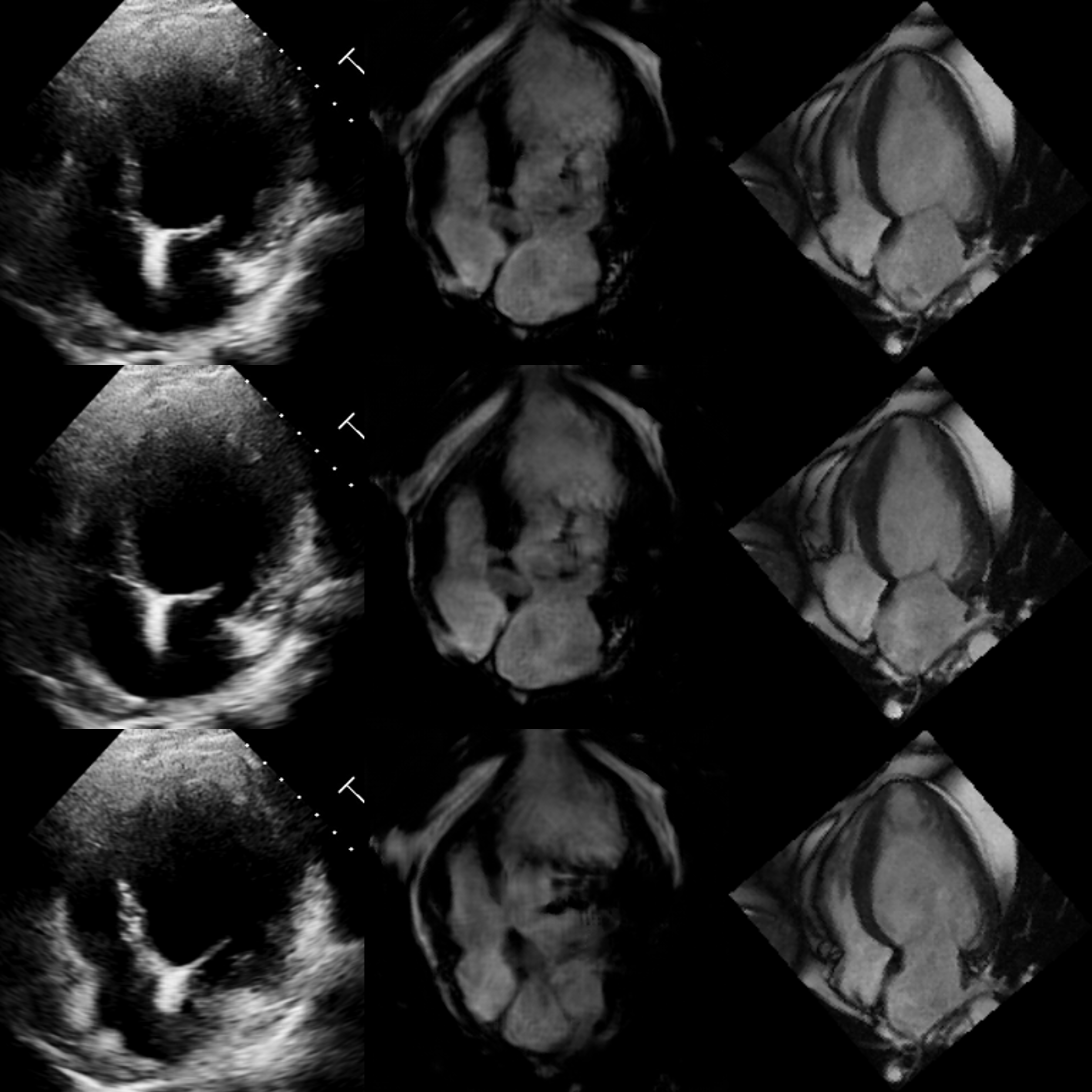}};
            \begin{scope}[x={(image.south east)}, y={(image.north west)}]
                \node[anchor=center, text=white] at (0.03, 0.97) {A};
                \node[anchor=center, text=white] at (0.37, 0.97) {B};
                \node[anchor=center, text=white] at (0.70, 0.97) {C};
            \end{scope}
        \end{tikzpicture}
        \captionof{figure}{\textbf{Patient 68:} Comparison of echocardiography (Column A), synthetic cardiac MRI (Column B), and real cardiac MRI (Column C) images. Echocardiography images show 3 different phases of one heartbeat, with corresponding synthetic and real cardiac MRI views.}
        \label{fig:AppendixPatient68}
    \end{minipage}
    \hfill
    \begin{minipage}[t]{0.49\textwidth}
        \centering
        \begin{tikzpicture}
            \node[anchor=south west, inner sep=0] (image) at (0,0) {%
              \includegraphics[width=\linewidth]{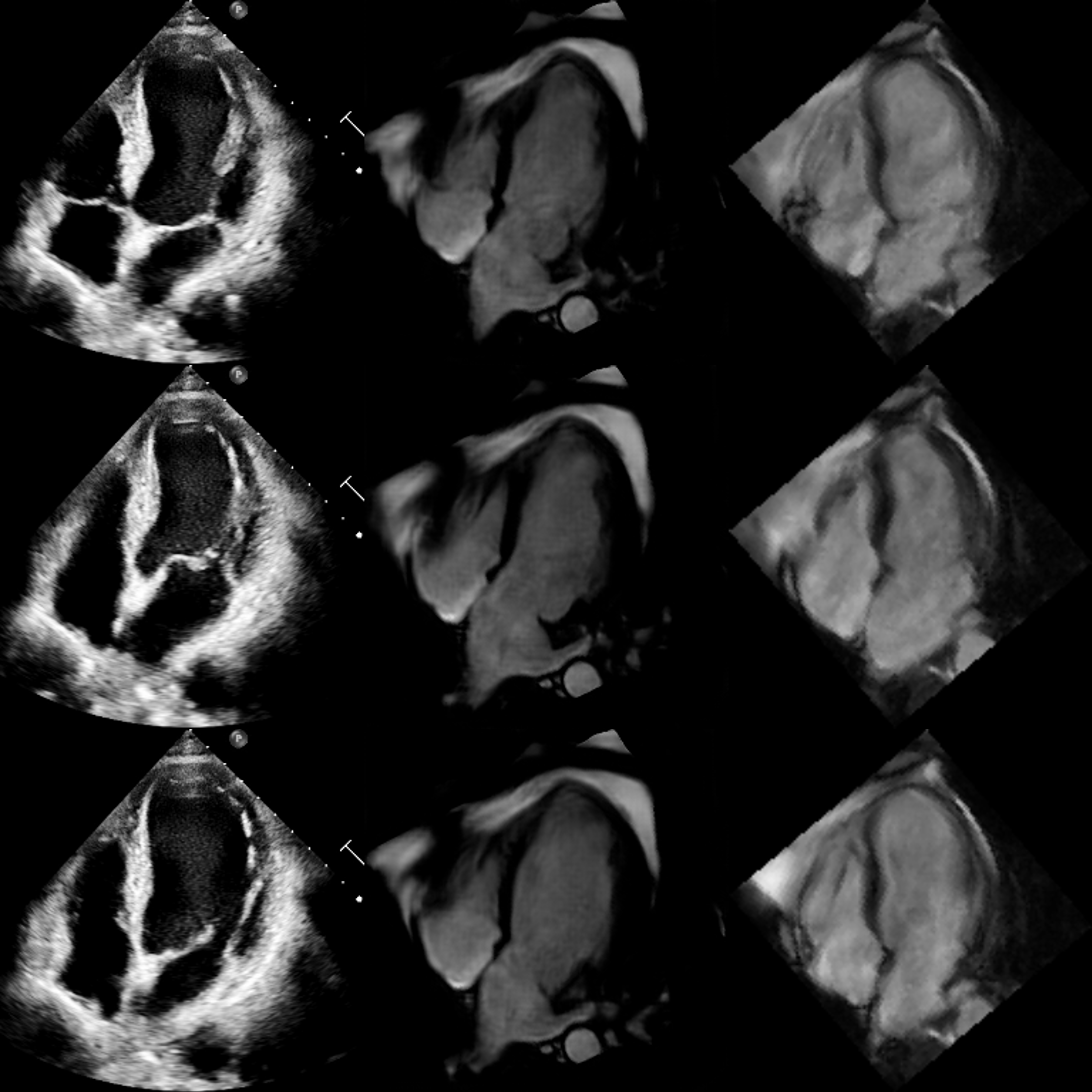}};
            \begin{scope}[x={(image.south east)}, y={(image.north west)}]
                \node[anchor=center, text=white] at (0.03, 0.97) {A};
                \node[anchor=center, text=white] at (0.37, 0.97) {B};
                \node[anchor=center, text=white] at (0.70, 0.97) {C};
            \end{scope}
        \end{tikzpicture}
        \captionof{figure}{\textbf{Patient 77:} Comparison of echocardiography (Column A), synthetic cardiac MRI (Column B), and real cardiac MRI (Column C) images. Echocardiography images show 3 different phases of one heartbeat, with corresponding synthetic and real cardiac MRI views.}
        \label{fig:AppendixPatient77}
    \end{minipage}
\end{figure}

\clearpage 
\newpage

\section*{\textbf{Appendix B}: Confusion Test Samples for Medical Evaluation}

\begin{figure}[!h]
    \centering
    \begin{minipage}[t]{0.48\textwidth}
    \centering
    \begin{tikzpicture}
        \node[anchor=south west, inner sep=0] (image) at (0,0) {
          \includegraphics[width=\linewidth]{adali9_7.png}
        };
        \begin{scope}[x={(image.south east)}, y={(image.north west)}]
            \node[anchor=north west, text=white] at (0.02,0.98) {A}; 
            \node[anchor=north west, text=white] at (0.52,0.98) {B}; 
            \node[anchor=north west, text=white] at (0.02,0.48) {C}; 
            \node[anchor=north west, text=white] at (0.52,0.48) {D}; 
        \end{scope}
    \end{tikzpicture}
    \captionof{figure}{Four samples from the confusion test, consisting of randomly selected synthetic and original cardiac MRI images. Cardiologists were asked to differentiate between the synthetic and original images. The samples include: A (original), B (original), C (synthetic), and D (synthetic).}
    \label{fig:Appendix2Page7}
    \end{minipage}
        \hfill
        \begin{minipage}[t]{0.48\textwidth}
        \centering
        \begin{tikzpicture}
            \node[anchor=south west, inner sep=0] (image) at (0,0) {
              \includegraphics[width=\linewidth]{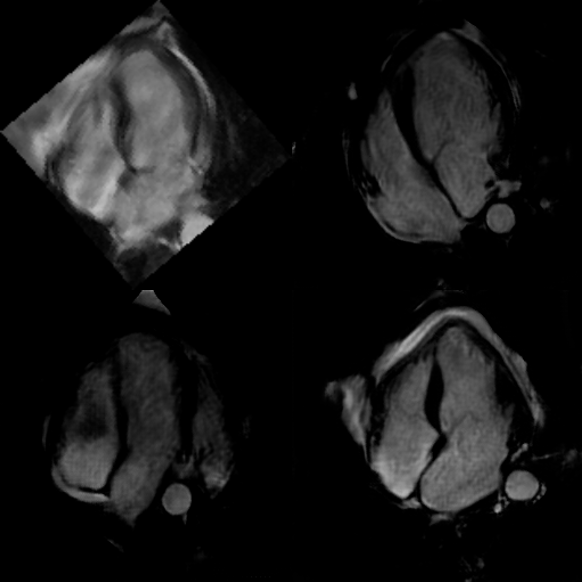}
            };
            \begin{scope}[x={(image.south east)}, y={(image.north west)}]
                \node[anchor=north west, text=white] at (0.02,0.98) {A};
                \node[anchor=north west, text=white] at (0.52,0.98) {B};
                \node[anchor=north west, text=white] at (0.02,0.48) {C};
                \node[anchor=north west, text=white] at (0.52,0.48) {D};
            \end{scope}
        \end{tikzpicture}
        \captionof{figure}{Four samples from the confusion test, consisting of randomly selected synthetic and original cardiac MRI images. Cardiologists were asked to differentiate between the synthetic and original images. The samples include: A (original), B (synthetic), C (synthetic), and D (synthetic).}
        \label{fig:Appendix2Page12}
    \end{minipage}
    
    \vspace{1em}
    
    \begin{minipage}[t]{0.48\textwidth}
    \centering
    \begin{tikzpicture}
        \node[anchor=south west, inner sep=0] (image) at (0,0) {
          \includegraphics[width=\linewidth]{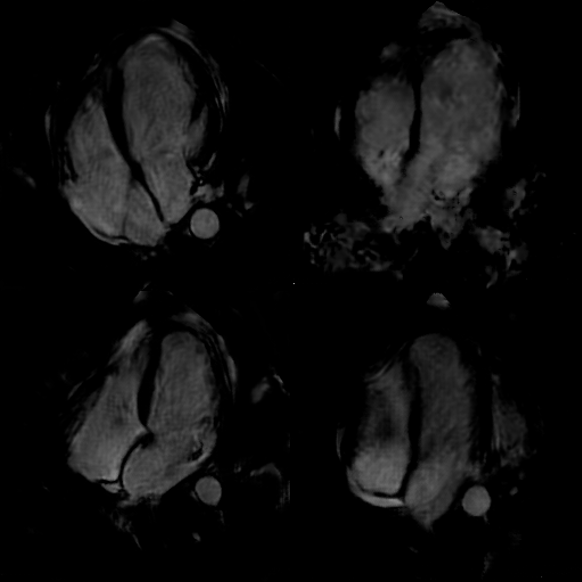}
        };
        \begin{scope}[x={(image.south east)}, y={(image.north west)}]
            \node[anchor=north west, text=white] at (0.02,0.98) {A};
            \node[anchor=north west, text=white] at (0.52,0.98) {B};
            \node[anchor=north west, text=white] at (0.02,0.48) {C};
            \node[anchor=north west, text=white] at (0.52,0.48) {D};
        \end{scope}
    \end{tikzpicture}
    \captionof{figure}{Four samples from the confusion test, consisting of randomly selected synthetic and original cardiac MRI images. Cardiologists were asked to differentiate between the synthetic and original images. The samples include: A (synthetic), B (synthetic), C (synthetic), and D (synthetic).}
    \label{fig:Appendix2Page28}
\end{minipage}
    \hfill
    \begin{minipage}[t]{0.48\textwidth}
    \centering
    \begin{tikzpicture}
        \node[anchor=south west, inner sep=0] (image) at (0,0) {
          \includegraphics[width=\linewidth]{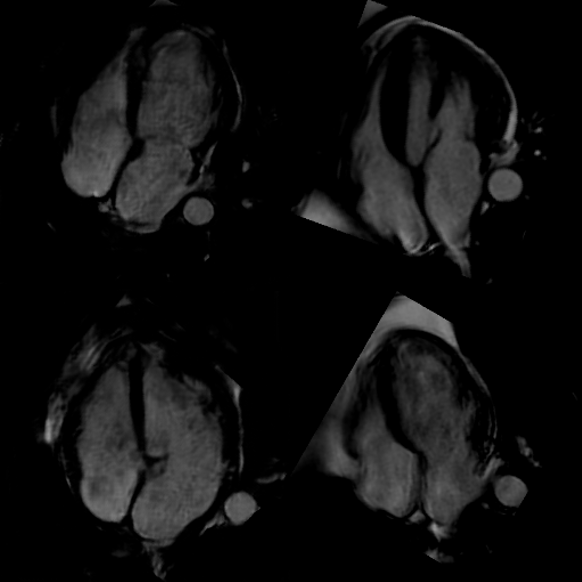}
        };
        \begin{scope}[x={(image.south east)}, y={(image.north west)}]
            \node[anchor=north west, text=white] at (0.02,0.98) {A};
            \node[anchor=north west, text=white] at (0.52,0.98) {B};
            \node[anchor=north west, text=white] at (0.02,0.48) {C};
            \node[anchor=north west, text=white] at (0.52,0.48) {D};
        \end{scope}
    \end{tikzpicture}
    \captionof{figure}{Four samples from the confusion test, consisting of randomly selected synthetic and original cardiac MRI images. Cardiologists were asked to differentiate between the synthetic and original images. The samples include: A (synthetic), B (original), C (synthetic), and D (original).}
    \label{fig:Appendix2Page31}
\end{minipage}
\end{figure}
\begin{figure}[!h]
    \centering
    \begin{minipage}[t]{0.48\textwidth}
    \centering
    \begin{tikzpicture}
        \node[anchor=south west, inner sep=0] (image) at (0,0) {
          \includegraphics[width=\linewidth]{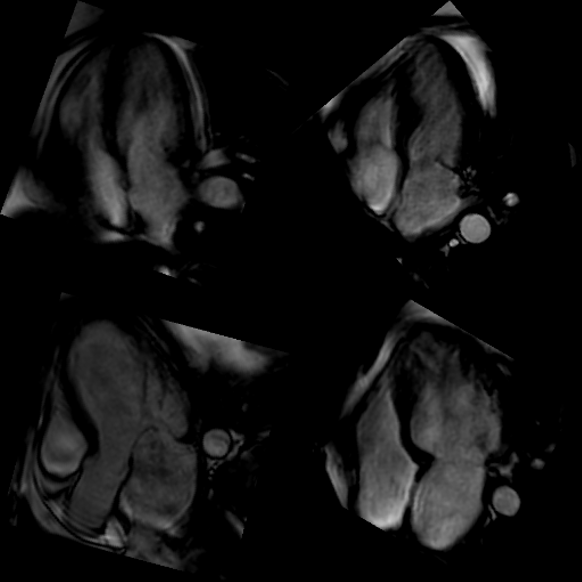}
        };
        \begin{scope}[x={(image.south east)}, y={(image.north west)}]
            \node[anchor=north west, text=white] at (0.02,0.98) {A}; 
            \node[anchor=north west, text=white] at (0.52,0.98) {B}; 
            \node[anchor=north west, text=white] at (0.02,0.48) {C}; 
            \node[anchor=north west, text=white] at (0.52,0.48) {D}; 
        \end{scope}
    \end{tikzpicture}
    \captionof{figure}{Four samples from the confusion test, consisting of randomly selected synthetic and original cardiac MRI images. Cardiologists were asked to differentiate between the synthetic and original images. The samples include: A (original), B (original), C (original), and D (original).}
    \label{fig:Appendix2Page33}
    \end{minipage}
        \hfill
        \begin{minipage}[t]{0.48\textwidth}
        \centering
        \begin{tikzpicture}
            \node[anchor=south west, inner sep=0] (image) at (0,0) {
              \includegraphics[width=\linewidth]{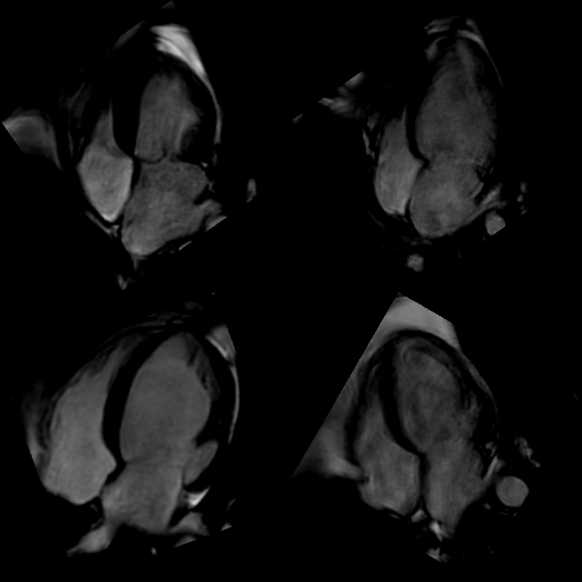}
            };
            \begin{scope}[x={(image.south east)}, y={(image.north west)}]
                \node[anchor=north west, text=white] at (0.02,0.98) {A};
                \node[anchor=north west, text=white] at (0.52,0.98) {B};
                \node[anchor=north west, text=white] at (0.02,0.48) {C};
                \node[anchor=north west, text=white] at (0.52,0.48) {D};
            \end{scope}
        \end{tikzpicture}
        \captionof{figure}{Four samples from the confusion test, consisting of randomly selected synthetic and original cardiac MRI images. Cardiologists were asked to differentiate between the synthetic and original images. The samples include: A (original), B (original), C (original), and D (original).}
        \label{fig:Appendix2Page35}
    \end{minipage}
    
    \vspace{1em}
    
    \begin{minipage}[t]{0.48\textwidth}
    \centering
    \begin{tikzpicture}
        \node[anchor=south west, inner sep=0] (image) at (0,0) {
          \includegraphics[width=\linewidth]{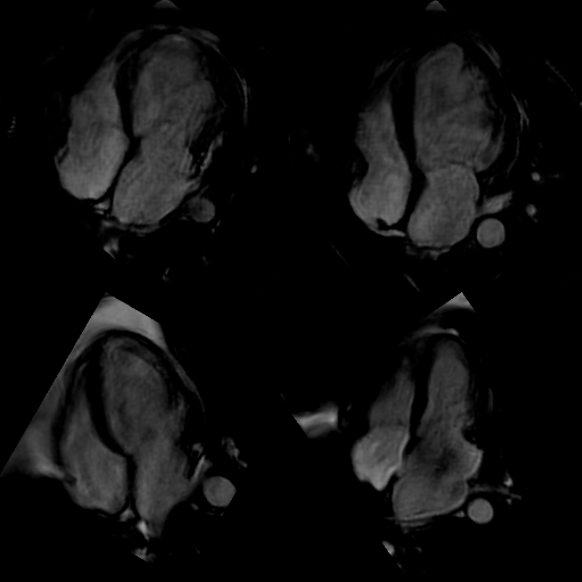}
        };
        \begin{scope}[x={(image.south east)}, y={(image.north west)}]
            \node[anchor=north west, text=white] at (0.02,0.98) {A};
            \node[anchor=north west, text=white] at (0.52,0.98) {B};
            \node[anchor=north west, text=white] at (0.02,0.48) {C};
            \node[anchor=north west, text=white] at (0.52,0.48) {D};
        \end{scope}
    \end{tikzpicture}
    \captionof{figure}{Four samples from the confusion test, consisting of randomly selected synthetic and original cardiac MRI images. Cardiologists were asked to differentiate between the synthetic and original images. The samples include: A (synthetic), B (synthetic), C (original), and D (original).}
    \label{fig:image36}
\end{minipage}
    \hfill
    \begin{minipage}[t]{0.48\textwidth}
    \centering
    \begin{tikzpicture}
        \node[anchor=south west, inner sep=0] (image) at (0,0) {
          \includegraphics[width=\linewidth]{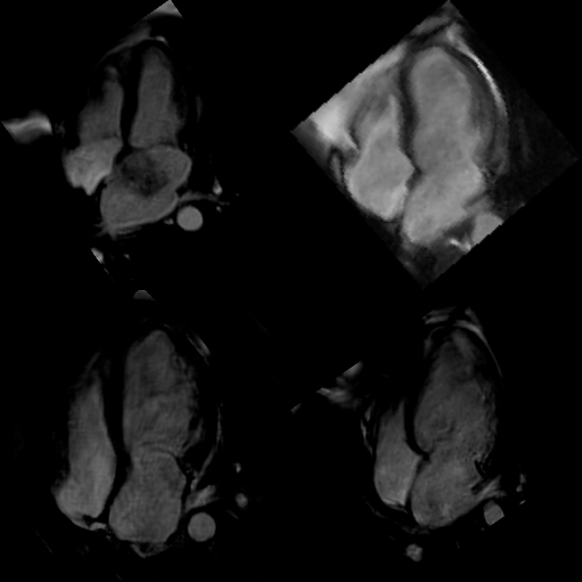}
        };
        \begin{scope}[x={(image.south east)}, y={(image.north west)}]
            \node[anchor=north west, text=white] at (0.02,0.98) {A};
            \node[anchor=north west, text=white] at (0.52,0.98) {B};
            \node[anchor=north west, text=white] at (0.02,0.48) {C};
            \node[anchor=north west, text=white] at (0.52,0.48) {D};
        \end{scope}
    \end{tikzpicture}
    \captionof{figure}{Four samples from the confusion test, consisting of randomly selected synthetic and original cardiac MRI images. Cardiologists were asked to differentiate between the synthetic and original images. The samples include: A (original), B (original), C (synthetic), and D (original).}
    \label{fig:Appendix2Page40}
\end{minipage}
\end{figure}
\begin{figure}[!h]
    \centering
    \begin{minipage}[t]{0.48\textwidth}
    \centering
    \begin{tikzpicture}
        \node[anchor=south west, inner sep=0] (image) at (0,0) {
          \includegraphics[width=\linewidth]{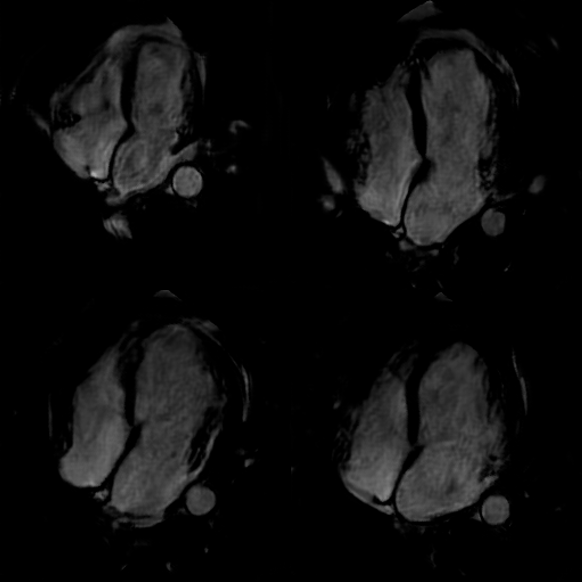}
        };
        \begin{scope}[x={(image.south east)}, y={(image.north west)}]
            \node[anchor=north west, text=white] at (0.02,0.98) {A}; 
            \node[anchor=north west, text=white] at (0.52,0.98) {B}; 
            \node[anchor=north west, text=white] at (0.02,0.48) {C}; 
            \node[anchor=north west, text=white] at (0.52,0.48) {D}; 
        \end{scope}
    \end{tikzpicture}
    \captionof{figure}{Four samples from the confusion test, consisting of randomly selected synthetic and original cardiac MRI images. Cardiologists were asked to differentiate between the synthetic and original images. The samples include: A (synthetic), B (synthetic), C (synthetic), and D (synthetic).}
    \label{fig:Appendix2Page45}
    \end{minipage}
        \hfill
        \begin{minipage}[t]{0.48\textwidth}
        \centering
        \begin{tikzpicture}
            \node[anchor=south west, inner sep=0] (image) at (0,0) {
              \includegraphics[width=\linewidth]{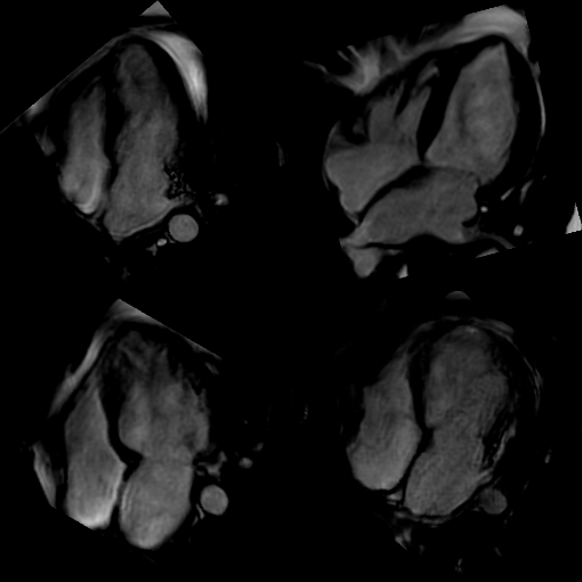}
            };
            \begin{scope}[x={(image.south east)}, y={(image.north west)}]
                \node[anchor=north west, text=white] at (0.02,0.98) {A};
                \node[anchor=north west, text=white] at (0.52,0.98) {B};
                \node[anchor=north west, text=white] at (0.02,0.48) {C};
                \node[anchor=north west, text=white] at (0.52,0.48) {D};
            \end{scope}
        \end{tikzpicture}
        \captionof{figure}{Four samples from the confusion test, consisting of randomly selected synthetic and original cardiac MRI images. Cardiologists were asked to differentiate between the synthetic and original images. The samples include: A (original), B (original), C (original), and D (synthetic).}
        \label{fig:Appendix2Page45}
    \end{minipage}
    
    \vspace{1em}
    
    \begin{minipage}[t]{0.48\textwidth}
    \centering
    \begin{tikzpicture}
        \node[anchor=south west, inner sep=0] (image) at (0,0) {
          \includegraphics[width=\linewidth]{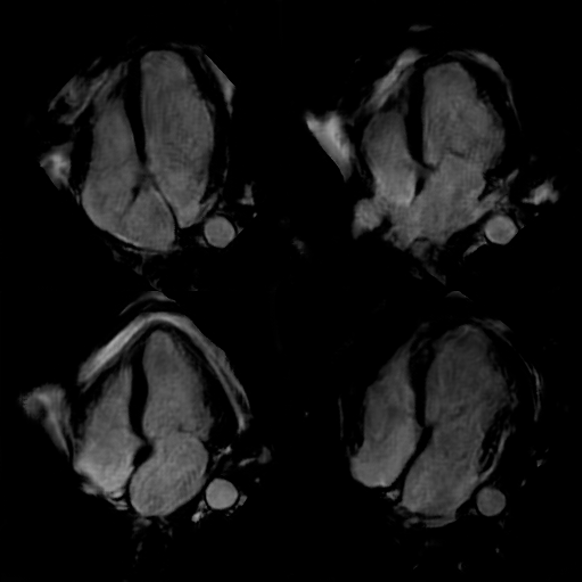}
        };
        \begin{scope}[x={(image.south east)}, y={(image.north west)}]
            \node[anchor=north west, text=white] at (0.02,0.98) {A};
            \node[anchor=north west, text=white] at (0.52,0.98) {B};
            \node[anchor=north west, text=white] at (0.02,0.48) {C};
            \node[anchor=north west, text=white] at (0.52,0.48) {D};
        \end{scope}
    \end{tikzpicture}
    \captionof{figure}{Four samples from the confusion test, consisting of randomly selected synthetic and original cardiac MRI images. Cardiologists were asked to differentiate between the synthetic and original images. The samples include: A (synthetic), B (synthetic), C (synthetic), and D (synthetic).}
    \label{fig:Appendix2Page49}
\end{minipage}
    \hfill
    \begin{minipage}[t]{0.48\textwidth}
    \centering
    \begin{tikzpicture}
        \node[anchor=south west, inner sep=0] (image) at (0,0) {
          \includegraphics[width=\linewidth]{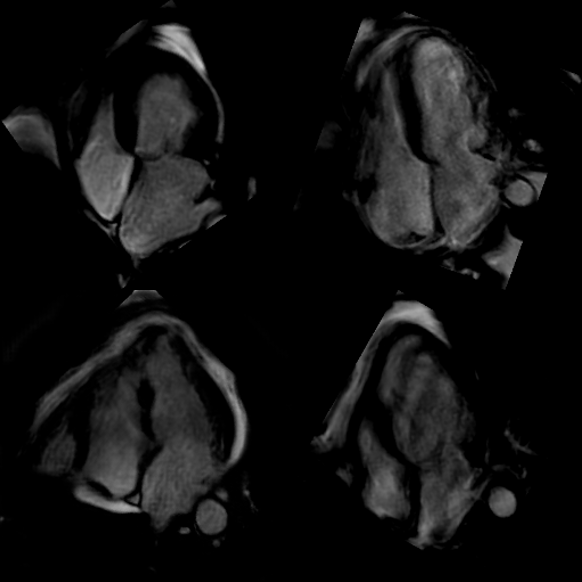}
        };
        \begin{scope}[x={(image.south east)}, y={(image.north west)}]
            \node[anchor=north west, text=white] at (0.02,0.98) {A};
            \node[anchor=north west, text=white] at (0.52,0.98) {B};
            \node[anchor=north west, text=white] at (0.02,0.48) {C};
            \node[anchor=north west, text=white] at (0.52,0.48) {D};
        \end{scope}
    \end{tikzpicture}
    \captionof{figure}{Four samples from the confusion test, consisting of randomly selected synthetic and original cardiac MRI images. Cardiologists were asked to differentiate between the synthetic and original images. The samples include: A (original), B (original), C (synthetic), and D (original).}
    \label{fig:Appendix2Page50}
\end{minipage}
\end{figure}
\end{document}